\documentclass[acmtog]{acmart}%,anonymous,review
\setcopyright{acmcopyright}
\copyrightyear{2026}
\acmYear{2026}
\acmDOI{XXXXXXX.XXXXXXX}

\acmJournal{TOG}
\acmVolume{0}
\acmNumber{0}
\acmArticle{0}
\acmMonth{0}

\usepackage[ruled,vlined]{algorithm2e}
\usepackage{makecell}
\usepackage[table]{xcolor}
\usepackage{arydshln}
\newtheorem{theorem}{Theorem}[section]
\newtheorem{definition}{Definition}[section]
\begin{document}
\title[MCHex: Marching Cubes Based Adaptive Hexahedral Mesh Generation with Guaranteed Positive Jacobian]{MCHex: Marching Cubes Based Adaptive Hexahedral Mesh Generation with Guaranteed Positive Jacobian}

%% Authors
\author{Hua Tong}
\email{huat2@andrew.cmu.edu}
\orcid{0009-0001-3010-2230}
\affiliation{
  \institution{Carnegie Mellon University}
  \city{Pittsburgh}
  \country{USA}
}

\author{Yongjie Jessica Zhang}
\email{jessicaz@andrew.cmu.edu}
\orcid{0000-0001-7436-9757}
\affiliation{
  \institution{Carnegie Mellon University}
  \city{Pittsburgh}
  \country{USA}}

\renewcommand{\shortauthors}{Tong et al.}

\begin{abstract}

Grid-based methods are the most robust approach for automatic hexahedral (hex) meshing, but they struggle to achieve high boundary fidelity and element quality. Conventional pipelines remove outside elements. This yields axis-aligned surfaces that converge to the input geometry at first order. The subsequent padding and projection steps are heuristic, offering no guarantees on final boundary fidelity or mesh quality. This paper introduces MCHex, a fundamental reformulation of boundary and mesh quality handling in grid-based hex meshing. MCHex directly applies Marching Cubes (MC) inside each grid cell, ensuring that the mesh boundary is an MC surface. The key insight is that by constraining cut-cell configurations to a 3-regular polyhedron, a midpoint subdivision of these configurations produces an all-hex mesh with a guaranteed positive Jacobian for every element. MCHex provides three advantages: (1) a theoretical guarantee of positive Jacobian for all hex elements; (2) boundary convergence that matches the approximation rate of MC, together with a non-heuristic algorithm that has well-bounded time complexity and achieves the fastest wall-clock time among all existing methods; and (3) generation of manifold surfaces for arbitrary geometries and a natural padded layer. Extensive evaluation on a benchmark of 202 geometries compares MCHex against several previous state-of-the-art hex meshing methods under a slightly smaller element budget, demonstrating that MCHex consistently produces positive Jacobian meshes with similar boundary fidelity while running significantly faster. MCHex can integrate seamlessly with post-processing steps such as mesh smoothing, mesh simplification, and is suitable for simulation.

\end{abstract}



%% Keywords

\begin{teaserfigure}
\centering
\includegraphics[width=\linewidth]{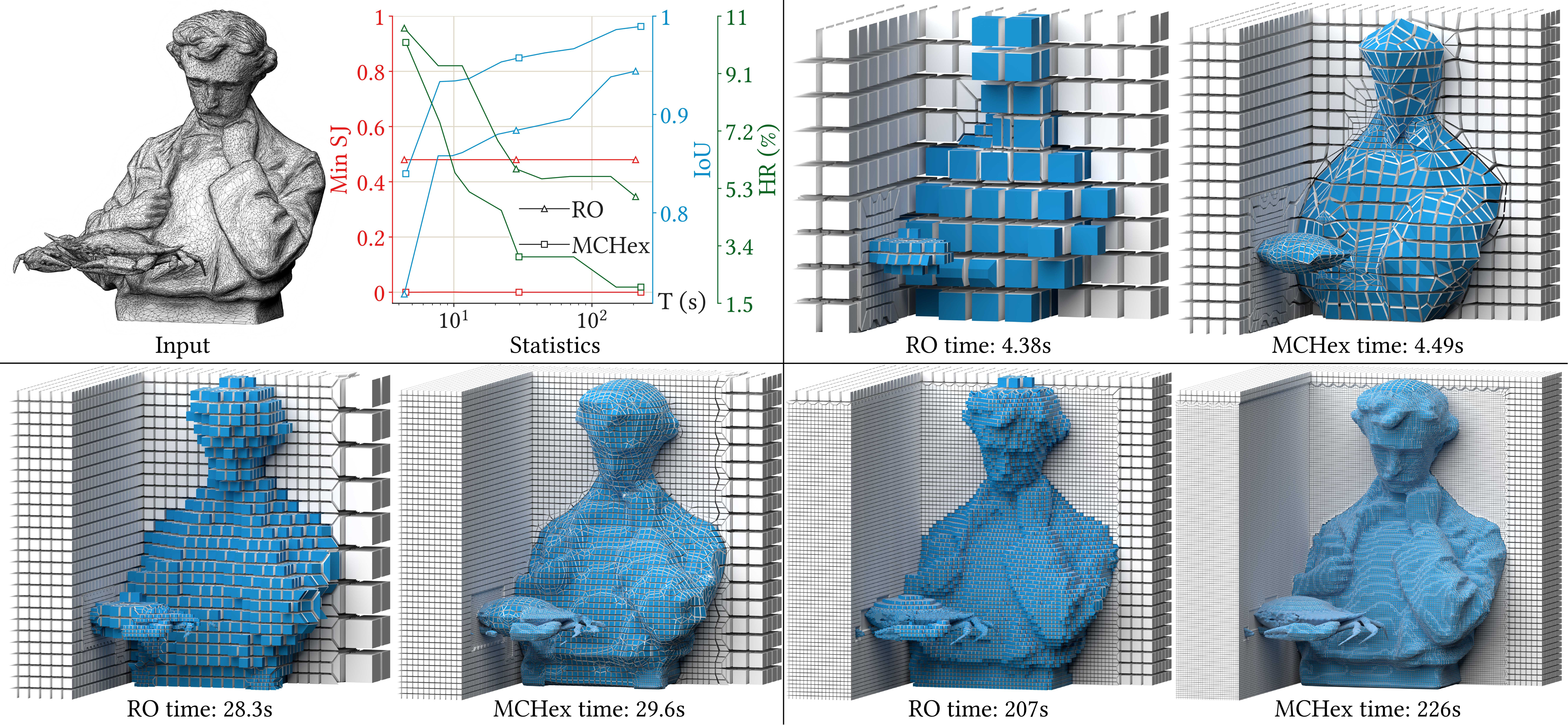}
\vspace{-8mm}
\Description{teaser}
\caption{Exhibition of 202 meshes hexahedralized by the novel hex meshing algorithm MCHex. The full dataset including the results from MCHex and previous methods is available at (the GitHub link will be provided upon acceptance).}
\label{fig:teaser}
\end{teaserfigure}

\maketitle
\section{Introduction}
\label{sec:introduction}

Hex mesh generation has long been regarded as a fundamental yet challenging problem in computational geometry and scientific computing. Hex meshes are highly desirable in finite element analysis (FEA) and isogeometric analysis \cite{zhang2016geometric} due to their superior numerical properties compared to tetrahedral (tet) meshes, offering better convergence rates, lower computational cost, and higher accuracy for many physical simulations \cite{cifuentes1992performance,benzley1995comparison,wang2004back,wang2021comparison}. Despite decades of research, automatically generating high-quality, all-hex meshes that approximate complex geometries while maintaining good mesh quality remains an open problem \cite{owen1998survey,blacker2000meeting,tautges2001generation,shepherd2008hexahedral,zhang2013challenges,schneider2022large}. Many different approaches have been explored, ranging from grid-based primal methods \cite{schneiders1996grid,tong2026element}, grid-based dual methods \cite{zhang2006adaptive,pitzalis2021generalized}, polycube map methods \cite{gregson2011all,mestrallet2025validity}, frame-field methods \cite{nieser2011cubecover,liu2023locally}, to machine learning methods \cite{yu2025dl,yu2026ddpm}; however, most of these methods lack robustness on general inputs, provide no quality guarantees, or take a long time to get the final mesh \cite{pietroni2022hex}.

Among various mesh generation strategies, the grid-based method has become the most reliable approach and the only one adopted in industrial practice. A prominent example is Hexotic \cite{marechal2016all}. The algorithm starts by adaptively refining a Cartesian grid until a specified criterion is met. Commonly used criteria include error-sensitive functions \cite{zhang2006adaptive,zhang2010automatic,zhang2013robust,hu2013adaptive}, normal similarity \cite{ito2009octree}, local thickness \cite{livesu2021optimal,pitzalis2021generalized,tong2024hybridoctree_hex,tong2026element}, and surface approximation accuracy \cite{gao2019feature,owen2017template}.

After refinement, hanging nodes in the grids are eliminated by replacing cells with transition templates that locally restore mesh conformity. Existing methods can be classified into primal and dual approaches. Primal methods directly incorporate hanging nodes into the output hex mesh. For 3-refinement schemes, the first method cannot handle concave regions effectively, leading to excessive refinement \cite{schneiders2000octree}. Subsequent works save more and more hex elements by developing templates for limited concave-edge configurations \cite{ito2009octree}, limited concave-corner configurations \cite{elsheikh2014consistent}, and eventually all transition configurations \cite{tong2026element}. In contrast, 2-refinement templates are more challenging to implement because they require pairing. However, the resulting meshes exhibit smoother mesh density transitions \cite{ebeida2011isotropic,zhang2013robust,owen2017template}.

Dual methods, which are exclusively applied to 2-refinement schemes, modify the input grids so that the dual mesh consists entirely of hex elements. These methods represent an important direction toward easier implementation and reduced element counts compared to primal methods. The first proposed method requires strong balancing and octree pairing \cite{marechal2009advances}, requirements that are retained in three subsequent pipelines \cite{marechal2016all,gao2019feature,tong2024hybridoctree_hex}. A later study optimizes the transition templates to reduce element counts \cite{hu2013adaptive}. A significant advancement is made by \cite{livesu2021optimal}, which introduces rotation-symmetric templates to mitigate irregular valence and expands the template set to relax the strong balancing condition to weak balancing and greatly reduces the element count. The most element-efficient method to date further relaxes octree pairing to generalized pairing by using linear programming to ensure an even number of same-density cells in the x-, y-, and z-directions \cite{pitzalis2021generalized}.

It is noteworthy that the reduction in element count comes at the expense of lowering the upper bound on mesh quality. For example, the original 3-refinement templates \cite{schneiders2000octree} yield a raw minimum scaled Jacobian (min SJ) of 0.480, whereas recent methods using weak balancing and generalized pairing produce min SJ on the order of \(10^{-3}\) \cite{livesu2021optimal,pitzalis2021generalized}. Although vertex-warping smoothing techniques can mitigate these issues, the increasingly complex topological connections inherently limit the upper bound of mesh quality that such smoothing can achieve.

The final step in the pipeline involves boundary projection. Due to the axis-aligned nature of grid-based meshing, boundary vertices must be projected onto the target geometry to accurately approximate the input surface. Naïvely projecting vertices directly results in poor element quality at the boundary. This degradation stems from a well-documented issue: when multiple faces of a single hex element lie on the boundary, projecting their vertices independently onto the target surface can induce significant distortion. Padding techniques address this issue by ensuring that each boundary element has only one face to be projected \cite{mitchell1995pillowing}. This has since become a standard step in all grid-based meshing pipelines. To ensure the quality of the padded elements, the surface vertex stars must be at least topologically homeomorphic to a disk \cite{ito2009octree,gao2019feature}, yet stricter conditions are sometimes required to achieve higher final mesh quality \cite{tong2024hybridoctree_hex}. Following padding, all methods apply smoothing techniques to project the padded vertices onto the target surface while improving mesh quality. Notably, while all preceding steps in the pipeline have deterministic time complexity with guaranteed mesh quality bounds, this projection step remains heuristic for state-of-the-art smoothing methods \cite{tong2025fast,tong2026hexopt,protais2026versatile}. Although there are earlier smoothing methods that ensure positive Jacobian \cite{freitag2000local,lin2015quality}, this guarantee is of limited practical value, as they do not provide a lower bound on the improvement of geometric fidelity between the initial axis-aligned boundary and the final smoothed boundary. The main observation of this paper is that this fundamental limitation is one of the key reasons for the slow progress in hex mesh generation over the past few decades. In contrast, tet mesh generation, backed by theoretical guarantees from constrained Delaunay tetrahedralization, both in terms of element quality and boundary preservation, has seen wider adoption in commercial applications.

This paper introduces MCHex, a novel hex meshing approach built upon the MC algorithm \cite{lorensen1998marching} and an improved variant \cite{kobbelt2001feature}. Unlike conventional pipelines that remove outside elements and produce axis-aligned boundaries that converge slowly to the input geometry, MCHex directly extracts the MC surface as the mesh boundary. This enables much faster geometric boundary convergence while rigorously guaranteeing positive Jacobians for all hex elements. MCHex is the fastest hex meshing tool in the benchmark evaluation.

In the pipeline, MCHex first adaptively refines a grid using the element-saving 3-refinement template to eliminate hanging nodes \cite{tong2026element}. Next, based on the Directed Distance Field (DDF) values computed on each edge relative to the input geometry \cite{kobbelt2001feature} (as the input triangle mesh is provided explicitly, DDF is more accurate than the Signed Distance Field), an MC surface is extracted within each hex element. For every grid cell, the reconstruction error between the input surface and the MC surface is estimated via sampling an error metric. MCHex employs Intersection over Union (IoU), as refining directly on the Hausdorff distance (HD) might bias the comparison when later evaluating against previous art that reports HD. Finally, to generate the hex volume mesh, it is observed that the reconstructed MC surface consistently partitions each cell into two or more polyhedra whose graph structures are necessarily 3-regular. Applying midpoint subdivision to a 3-regular polyhedron yields a Closure-finite Weak topology complex (CW complex) composed entirely of hex elements. Each resulting hex element shares at most one face with the MC surface, which ensures automatic padding. Another key insight is that by offsetting interior points and, in certain special cases, clamping the DDF values, the Jacobian of every hex element is guaranteed to be positive. As a result, the MCHex method offers the following three key advantages:

\begin{enumerate}
\item The Jacobian of every hex element is guaranteed to be positive with rigorous proof.
\item It converges significantly faster to the input geometry than previous grid-based hex meshing pipelines that remove outside elements and project the zigzag boundary to the input geometry. The algorithm is non-heuristic, with runtime linear in the number of hex elements (\#hex) and sublinear in the number of input triangles (\#tri), and peak RAM linear in both \#hex and \#tri.
\item MC guarantees a manifold surface with a naturally padded boundary, and a local refinement lookup table analogous to the 31 MC configurations \cite{chernyaev1995marching} mitigates topology ambiguities within each cell.
\end{enumerate}

In the experiments, MCHex is tested on the benchmark released by \cite{gao2019feature} comprising 202 input geometries. The resulting meshes are shown in Figure \ref{fig:teaser}. A large-scale comparison is conducted with selected prior hex meshing art, evaluating runtime, IoU, Hausdorff ratio (HR) which is the HD normalized by the bounding box diagonal, and five mesh quality metrics. The results demonstrate that using the fewest hex elements, MCHex achieves the fastest generation speed, and the geometry fidelity and mesh quality are comparable to state-of-the-art methods.

The remainder of the paper is organized as follows. Section \ref{sec:rationale} presents the rationale of the MCHex method. Section \ref{sec:algorithmOverview} provides an overview of the proposed algorithm. Section \ref{sec:propertiesof3RegularGraphs} outlines the beneficial properties of 3-regular graphs that guarantee the generation of all-hex meshes following midpoint subdivision. Section \ref{sec:positiveJacobianProof} provides a proof that all hexes have positive Jacobian. Section \ref{sec:resultsandApplications} provides an extensive evaluation on the benchmark \cite{gao2019feature}. Finally, Section \ref{sec:conclusionandFutureWork} summarizes the key contributions and suggests potential future research directions. For full reproducibility, the reference implementation is provided in the supplementary materials and will be made publicly available on GitHub after the paper is accepted.

\section{Rationale}
\label{sec:rationale}

\begin{figure*}
\centering
\includegraphics[width=\linewidth]{comparePreviousMethodsWithMCHex.jpg}
\vspace{-8mm}
\Description{comparePreviousMethodsWithMCHex}
\caption{Adapting a uniform grid (black) to an input curve (red) using the baseline removing outside element method, midpoint MC \cite{protais2025automatic}, and MCHex. Black dots are points inside the input geometry. The baseline extracts an axis-aligned boundary that converges to the input geometry at first order. Projection (a heuristic step) creates an inverted element (white), necessitating padding to split the reflex angle. In \cite{protais2025automatic}, the boundary is extracted with an MC-like algorithm, whereas it uses the midpoint of sign-change edges instead of the true intersection to preserve good Jacobians. Its boundary also converges at first order. MCHex uses the true intersection, giving a boundary that converges at second order and lies close to the input geometry even without projection.}
\label{fig:comparePreviousMethodsWithMCHex}
\end{figure*}

The foundational concept for the MCHex approach is inspired by the work on adaptive quadrilateral (quad) meshing without cleanup operations \cite{rushdi2017all} and the work on uniform hex meshing using Dhondt's cut algorithm \cite{protais2025automatic}, as summarized in Figure \ref{fig:comparePreviousMethodsWithMCHex}. The left three panels illustrate the baseline commonly employed in previous grid-based hex/quad meshing pipelines. This method exhibits two significant limitations: (1) the resulting boundary is axis-aligned. The normal never converges to the input geometry, and the HD converges to the input geometry at only first order, and (2) as a consequence of this poor initial approximation, subsequent projection induces distortion in boundary elements, making them susceptible to inversion (an inverted element is shown in white). To mitigate such inversions, additional padding of boundary elements is required to split reflex angles. In contrast, the middle two panels show one MC-based approach \cite{protais2025automatic}. It adopts an MC-like algorithm, but takes the midpoint of sign-change edges instead of the true intersection. Using midpoints guarantees good positive Jacobian values for all generated elements (it is also easy to check because all hex element shapes form a finite set that can be enumerated). Its boundary converges to the input geometry at first order. The right panel shows the proposed MCHex algorithm, which uses the true intersection. The planar quad scenario is studied in \cite{rushdi2017all}. As a result, its boundary converges to the input geometry at second order (a simplified proof of uniform refinement scenario is in Appendix \ref{apd:boundaryApproximationConvergenceRate}) and, before projection, already lies very close to the input geometry. Regarding mesh quality, in \cite{rushdi2017all}, poorly shaped elements are addressed by implementing a vertex repelling strategy: when intersection points approach too closely to grid points, the grid points are repelled to maintain reasonable edge ratios and Jacobians. Thanks to the direct geometric interpretation of quad Jacobian as the sine value of the interior angle, this procedure theoretically guarantees positive Jacobian quality.

However, extending this approach to three dimensions introduces substantial challenges. The authors of \cite{rushdi2017all} also propose a three-dimensional hex meshing variant \cite{awad2016all}, but it fails to recognize that the essence lies in combining volumetric meshing with MC. Instead, the authors simplify the problem to five basic cases of planar-cube intersections. In practice, the topological configurations of input geometry within a cube are more complex, and intersection points on cutting surfaces with more than three vertices are typically non-coplanar. As a result, the demonstrated results are restricted to simple geometries: a sphere, a doll, and a cylinder, without providing guarantees on mesh quality. min SJ of \(0.277\) is guaranteed before projection in \cite{protais2025automatic}, whereas the boundary still converges at first order. In addition, the provided configurations cannot solve the problem of face and volume topology ambiguities. Furthermore, both methods are restricted to uniform grids, unlike the adaptive grid framework employed in the two-dimensional scenario \cite{rushdi2017all}. The natural next step is therefore to investigate whether all polyhedra that arise in MC can be subdivided into all-hex elements, and whether mesh quality guarantees can be maintained under adaptive background grids with intersection points moving on edges. Here, two primary difficulties emerge. On one hand, while subdividing an arbitrary \(n\)-gon in two dimensions invariably produces \(n\) quads, the three-dimensional case imposes stricter constraints. As discussed later in Section \ref{sec:propertiesof3RegularGraphs}, only polyhedra with 3-regular graph connectivity containing \(n\) vertices are guaranteed to yield \(n\) hexes after subdivision. On the other hand, the geometric interpretation of Jacobian quality becomes significantly more complex. In two dimensions, the Jacobian is related to the sine of the interior angles; in three dimensions, it corresponds to the signed volume of the tetrahedron formed by a vertex's three edges. This complexity, together with tremendous configurations in adaptive grids, makes establishing rigorous quality guarantees significantly more difficult. This paper aims to address these two challenges by providing theoretical foundations and practical solutions, as a step toward robust hex meshing.

\section{Algorithm Overview}
\label{sec:algorithmOverview}

\begin{figure}
\centering
\includegraphics[width=\linewidth]{overview.jpg}
\vspace{-8mm}
\Description{overview}
\caption{The basic steps of MCHex. Top row (left to right): (1)-(3); bottom row (left to right): (4)-(5). (1) Input geometry \(T_\text{geom}\) (red) and AABB \(T_\text{AABB}\) (black), (2) adaptive grid \(H_{\text{Int}(T_\text{AABB})}\) obtained by iteratively refining cells until the misclassified volume in each cell falls below a user-provided threshold \(\epsilon_{\text{vol}}\), (3) conforming hex mesh \(H'_{\text{Int}(T_\text{AABB})}\) after removal of hanging nodes \cite{tong2026element}, (4) polyhedra (blue) produced by MC, and (5) hex mesh produced by midpoint subdivision for the interior \(H_{\text{Int}(T_\text{geom})}\) (blue) and the exterior \(H_{\text{Int}(T_\text{AABB})\setminus\text{Int}(T_\text{geom})}\). Notice that sharp features are smoothed out due to the limitation of MC.}
\label{fig:overview}
\end{figure}

Of particular interest in the storage of three-dimensional geometry is a closed pure two-dimensional simplicial complex \(T_\text{geom}\). According to the Jordan-Brouwer separation theorem \cite{hatcher2001algebraic}, \(T_\text{geom}\) partitions \( \mathbb{R}^3 \) into two connected components: a bounded internal region \( \text{Int}(T_\text{geom}) \) and an unbounded external region \( \mathbb{R}^3\setminus\text{Int}(T_\text{geom}) \). To obtain a finite computational domain, an axis-aligned bounding box (AABB) \( T_\text{AABB} \) is introduced, which also partitions \( \mathbb{R}^3 \) into a bounded internal region \( \text{Int}(T_\text{AABB}) \) and an unbounded external region \( \mathbb{R}^3\setminus\text{Int}(T_\text{AABB}) \), with the condition that \( \text{Int}(T_\text{geom}) \subset \text{Int}(T_\text{AABB}) \).

Let \( M \subset \mathbb{R}^3 \) be a solid region. The hex mesh \( H_M = (V, E, Q, H) \) is a CW complex \cite{hatcher2001algebraic} of hex elements \( H \) that decomposes \( M \). \(V,E,Q\) and \(H\) represent the vertices, the edges, the quad faces, and the hex elements, respectively. The boundary \( \partial H \) consists of those faces in \( Q \) belonging to only one hex element, along with their constituent edges and vertices. Hex elements in \( H \) that are not part of \( \partial H \) are referred to as internal elements. The domain \(\text{Int}(T_{\text{AABB}})\) is tessellated into a hex mesh \( H_{\text{Int}(T_\text{AABB})} \). Given the hex mesh \(H_{\text{Int}(T_{\text{AABB}})}\), the algorithm extracts \(H_{\text{Int}(T_\text{geom})}\) and \(H_{\text{Int}(T_\text{AABB})\setminus\text{Int}(T_\text{geom})}\), both of which are guaranteed to satisfy the following properties: positive Jacobian, accurate boundary (i.e., second-order convergence rate of HR), and \(C^0\)-conforming along their shared boundary. ``\(C^0\)-conforming'' means two piecewise linear meshes are topologically and geometrically consistent across their common quad boundary faces. In detail, there exists a bijection \(\phi: \partial H_{\text{Int}(T_\text{geom})} \to \partial H_{\text{Int}(T_\text{AABB})\setminus\text{Int}(T_\text{geom})}\) such that for each quad face \(q(V_q, E_q) \in \partial H_{\text{Int}(T_\text{geom})}\), face \(\phi(q) \in \partial H_{\text{Int}(T_\text{AABB})\setminus\text{Int}(T_\text{geom})}\) satisfies \(\phi(v) = v\) and \(\phi(e) = e\) for \(v \in V_q\) and \(e \in E_q\).

\begin{algorithm}
\caption{Hex meshing of an input geometry\label{alg:entirePipeline}}
\KwIn{Closed pure two-dimensional simplicial complex \(T_\text{geom}\), user-provided misclassified volume threshold \(\epsilon_\text{vol}\)}
\KwOut{Hex meshes \(H_{\text{Int}(T_\text{geom})}\) and \(H_{\text{Int}(T_\text{AABB})\setminus\text{Int}(T_\text{geom})}\)}
\hrule
Initialize axis-aligned bounding box grid \(H_{\text{Int}(T_{\text{AABB}})}\)\;
Mark the root cell as needing refinement\;
\While{some cells in \(H_{\text{Int}(T_{\text{AABB}})}\) need refinement}{
    \For{each cell in \(H_{\text{Int}(T_{\text{AABB}})}\)}{
        \If{the cell needs refinement}{
            Refine cell\;
            Compute and clamp DDF values for new edges in child cells to \([\epsilon,1-\epsilon]\) \cite{kobbelt2001feature}\;
        }
    }
    Extract \(C^0\)-conforming hex mesh \(H'_{\text{Int}(T_{\text{AABB}})}\) by removing hanging nodes in \(H_{\text{Int}(T_{\text{AABB}})}\) \cite{tong2026element}\;
    Compute and clamp DDF values for new edges in \(H'_{\text{Int}(T_{\text{AABB}})}\)\;
    Determine MC topology configurations by solving 31 face and volume topology ambiguities in \(H'_{\text{Int}(T_{\text{AABB}})}\) \cite{chernyaev1995marching}\;
    Mark all cells in \(H_{\text{Int}(T_{\text{AABB}})}\) as not needing refinement\;
    \For{each cell in \(H_{\text{Int}(T_{\text{AABB}})}\)}{
        \If{sign-change edges exist}{
            Determine MC topology case in all hexes of \(H'_{\text{Int}(T_{\text{AABB}})}\) inside the cell\;
            \If{topology is one of the four *-ed complex cases}{
                \If{the cell is not a unit cube}{
                    Mark the cell as needing refinement\;
                    \KwSty{continue}
                }
                Apply local refinement template to subdivide the cell using ray casting for four *-ed cases\;
                Compute and clamp DDF values for new edges in the refined mesh\;
                Reconstruct surface on the refined mesh using MC\;
            }
            \Else{
                Reconstruct surface directly using MC\;
            }
            \If{misclassified volume in the cell \(>\epsilon_\text{vol}\)}{
                Mark the cell as needing refinement\;
            }
        }
    }
}
\For{each hex in \(H'_{\text{Int}(T_{\text{AABB}})}\)}{
    Generate hexes with MC and midpoint subdivision\;
    Warp volume center points and clamp intersection points based on a lookup table to ensure positive Jacobians\;
}
\Return \(H_{\text{Int}(T_\text{geom})}\) and \(H_{\text{Int}(T_\text{AABB})\setminus\text{Int}(T_\text{geom})}\)\;
\end{algorithm}

Algorithm \ref{alg:entirePipeline} summarizes the complete pipeline. Figure \ref{fig:overview} illustrates the geometric intuition in two dimensions. The process begins by initializing an axis-aligned bounding box. Following the same spirit as \cite{gao2019feature}, MCHex iteratively refines cells based on both geometric approximation errors and topological complexity. Due to the presence of topology ambiguities and hanging nodes, each refinement iteration must first extract the \(C^0\)-conforming hex mesh by removing hanging nodes, then determine MC topology configurations by resolving 31 face and volume topology ambiguities, after which the surface can be reconstructed inside each cell. For every cell that exhibits sign-change edges, its MC topology case is identified. If the topology is one of the four *-ed complex cases, the cell is marked for refinement unless it is already a unit cube, in which case a local refinement template using ray casting for four specific *-ed cases is applied, DDF values are computed and clamped, and the surface is reconstructed using MC. For base cases, the surface is reconstructed directly using MC. If the misclassified volume in the cell exceeds the user-provided threshold \(\epsilon_\text{vol}\), the cell is also marked for refinement in the next round. The misclassified volume is used because its computation can be done per cell. Other metrics, such as HR employed in \cite{gao2019feature}, might be more accurate, but they are often computed globally and therefore incur higher time complexity. Adopting HR as the refinement metric can be a future direction. Once no cells require further refinement, the final hex mesh is generated by executing MC using the clamped DDF values stored on the edges, followed by a midpoint subdivision step. To ensure positive Jacobian, volume center points are warped, and intersection points in challenging configurations are clamped based on theoretically derived values, as detailed later in Section \ref{sec:positiveJacobianProof}.

\section{Properties of 3-regular graphs}
\label{sec:propertiesof3RegularGraphs}

\begin{figure}
\centering
\includegraphics[width=\linewidth]{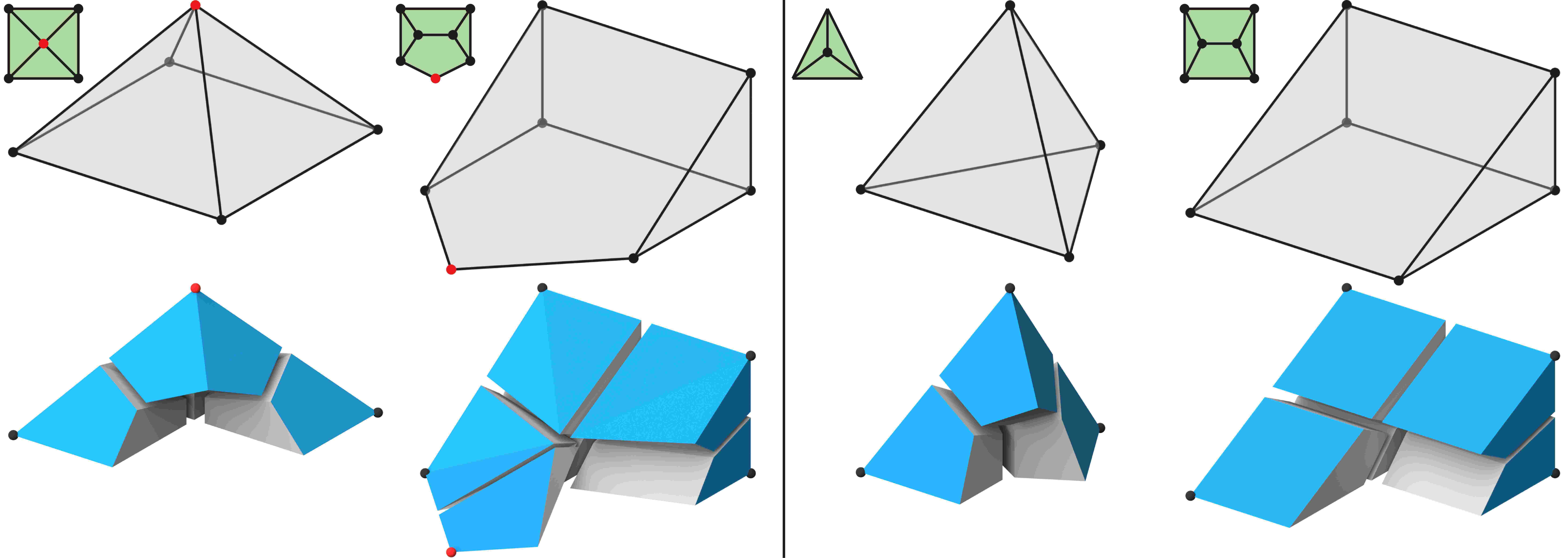}
\vspace{-8mm}
\Description{kRegularGraphMidpointSubdivision}
\caption{The left two polyhedra are non-3-regular, with violating vertices marked by red dots; the right two polyhedra are 3-regular. The top row illustrates the two-dimensional CW complexes (white) with 1-skeleton (green), while the bottom row shows the resulting three-dimensional volumetric meshes after applying midpoint subdivision. One element is removed in each example to show the interior.}
\label{fig:kRegularGraphMidpointSubdivision}
\end{figure}

This section begins by establishing the definitions of \(k\)-regular graph, \(k\)-regular polyhedron, and midpoint subdivision operation. It then provides a theorem that the midpoint subdivision of a polyhedron induces all-hex elements if and only if the polyhedron is 3-regular. The proof of this theorem is provided in Appendix \ref{apd:only3RegularPolyhedronAfterMidpointSubdivisionGivesAllHex}.

\begin{definition}[\(k\)-regular graph]
\label{def:kRegularGraph}
A \(k\)-regular graph is an undirected graph \(G = (V, E)\) where every vertex \(v \in V\) is incident to \(k\) edges or has \(k\) neighboring vertices.
\end{definition}

\begin{definition}[\(k\)-regular polyhedron]
\label{def:kRegularPolyhedron}
\(P = (V, E, F)\) is a closed pure two-dimensional CW complex, where each face in \(F\) is attached to a cycle in the 1-skeleton of \(P\). The 1-skeleton, which consists of the vertices \(V\) and edges \(E\), forms the underlying graph structure of the complex. If the underlying graph \((V, E)\) is a \(k\)-regular graph (where each vertex has \(k\) neighboring vertices), then \(P\) is a \(k\)-regular polyhedron.
\end{definition}
In the first row of Figure \ref{fig:kRegularGraphMidpointSubdivision}, the left two are non-3-regular polyhedra, while the right two are 3-regular polyhedra. Their 1-skeletons are shown at the top-left corner of each polyhedron in green.

\begin{figure}
\centering
\includegraphics[width=\linewidth]{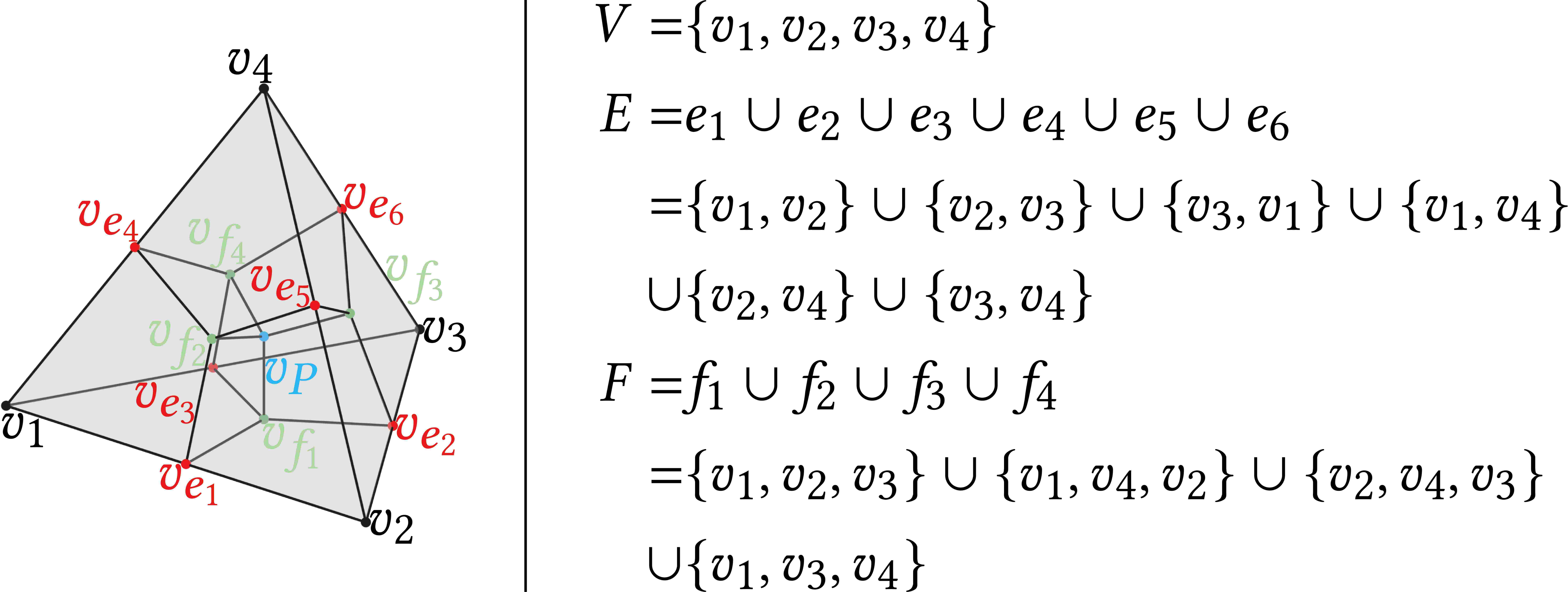}
\vspace{-8mm}
\Description{midpointSubdivisionExample}
\caption{An example of midpoint subdivision converting a tet boundary \(P=(V,E,F)\) into mesh \(P'=(V',E',F',C')\) composed of four hex elements, illustrating Theorem \ref{thm:only3RegularPolyhedronAfterMidpointSubdivisionGivesAllHex}.}
\label{fig:midpointSubdivisionExample}
\end{figure}

\begin{definition}[midpoint subdivision]
\label{def:midpointSubdivision}
The midpoint subdivision of a closed pure two-dimensional CW complex \(P = (V, E, F)\) induces a pure three-dimensional CW complex \(P'=(V',E',F',C')\) as follows: For each edge \(e(V_e)\in E\), insert edge midpoint \(v_e=\frac{1}{\lvert V_e\rvert}\sum_{v\in V_e}v\). For each face \(f(V_f, E_f) \in F\), insert face midpoint \(v_f=\frac{1}{\lvert V_f\rvert}\sum_{v\in V_f}v\), then for each edge \(e\in E_f\), connect edge midpoint \(v_e\) to \(v_f\). Insert volume center \(v_P=\frac{1}{\lvert V\rvert}\sum_{v\in V}v\). For each face \(f\in F\), connect the face midpoint \(v_f\) to \(v_P\). The resulting structure is induced to form \(P'=(V',E',F',C')\).
\end{definition}
The focus of this definition is the topological changes to the graph connectivity and the cellular complex by the subdivision. Geometrically, the positions of newly inserted points are usually midpoints (e.g., center points), as shown in the second row of Figure \ref{fig:kRegularGraphMidpointSubdivision} and in Figure \ref{fig:midpointSubdivisionExample}. However, as will be shown in Section \ref{sec:positiveJacobianProof}, using center positions sometimes results in negative Jacobians.

\begin{theorem}
\label{thm:only3RegularPolyhedronAfterMidpointSubdivisionGivesAllHex}
Let \(P=(V,E,F)\) be a closed pure two-dimensional CW complex, and \(P'=(V',E',F',C')\) be the pure three-dimensional CW complex induced by applying midpoint subdivision to \(P\). Then every element \(c\in C'\) is a hex if and only if \(P\) is a 3-regular polyhedron.
\end{theorem}
A natural next step is to investigate whether a sign-change cell produced by MC can always be decomposed into a set of 3-regular polyhedra. If this holds, then Theorem \ref{thm:only3RegularPolyhedronAfterMidpointSubdivisionGivesAllHex} guarantees that applying midpoint subdivision to each such polyhedron yields all hex elements. The remaining geometric problem is to ensure that every hex element Jacobian is positive.

\section{Positive Jacobian Proof}
\label{sec:positiveJacobianProof}

\begin{figure}
\centering
\includegraphics[width=\linewidth]{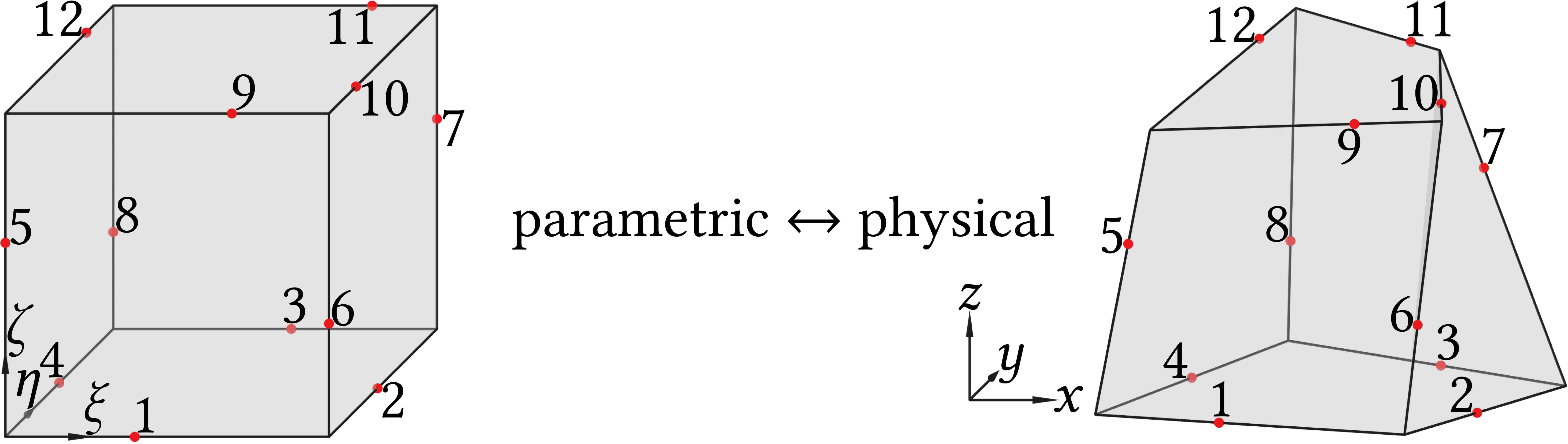}
\vspace{-8mm}
\Description{jacobian}
\caption{Coordinate transformation of a hex element from its local parametric space (left) to the global physical space (right). Twelve red points, numbered 1 through 12, represent the intersection points on twelve edges.}
\label{fig:jacobian}
\end{figure}

This section begins by defining the Jacobian as a key measure of element quality. MC with 31 topology ambiguity cases \cite{chernyaev1995marching} is then applied to reconstruct the surface within each cell \(h \in H_{\text{Int}(T_\text{AABB})}\) where the DDF values at its edges exhibit sign changes. It is shown that decomposed polyhedra are all 3-regular, making them hex-meshable via midpoint subdivision. Among a total of 31 topological cases in a cell, 14 templates that are guaranteed to generate positive Jacobian hexes under midpoint subdivision are listed and proved. For the remaining 17 complex cases where a strict positive Jacobian guarantee is difficult to establish, local refinement templates can theoretically be introduced to convert them into the 14 base cases. Notice that the positive Jacobian proofs are given for all 200 cell types in \cite{tong2026element}, not just for the unit cube case.

As shown in Figure \ref{fig:jacobian}, the finite element discretization of complex geometries employs a coordinate transformation to map hex elements with regular geometry in local coordinates \((\xi, \eta, \zeta)\) (parametric domain) in the left panel to distorted hex elements in global Cartesian coordinates \((x, y, z)\) (physical domain) in the right panel. This transformation is typically implemented using an interpolation scheme based on nodal shape functions, written as:
\begin{align}
\label{equ:xyzToxietazetaInterpolation}
\begin{bmatrix}
x,y,z
\end{bmatrix}^\top
= \mathbf{X}\cdot\mathbf{N}(\xi,\eta,\zeta),
\end{align}
where \(\mathbf{X}\) are the global coordinates of the eight vertices defining the hex. Usually, Lagrange polynomials \(\mathbf{N}\) are shape functions for the parameter domain with \(\xi, \eta, \zeta \in [0, 1]\), written as
\begin{align}
\label{equ:XN}
\mathbf{X} = 
\begin{bmatrix}
x_1,\dots,x_8 \\
y_1,\dots,y_8 \\
z_1,\dots,z_8
\end{bmatrix},
\mathbf{N}(\xi,\eta,\zeta)=
\begin{bmatrix}
(1-\xi)(1-\eta)(1-\zeta)\\\xi(1-\eta)(1-\zeta)\\\xi\eta(1-\zeta)\\(1-\xi)\eta(1-\zeta)\\(1-\xi)(1-\eta)\zeta\\\xi(1-\eta)\zeta\\\xi\eta\zeta\\(1-\xi)\eta\zeta
\end{bmatrix}.
\end{align}
These functions ensure \(C^0\)-continuity across adjacent elements and enable parametric representation of the distorted element geometry.

By differentiating Equation \eqref{equ:xyzToxietazetaInterpolation}, the Jacobian matrix \(\mathbf{J}\), which defines the local-to-global coordinate transformation, is
\begin{align}
\label{equ:jacobianDefinition}
\mathbf{J} = 
\begin{bmatrix}
\frac{\partial x}{\partial \xi},\frac{\partial x}{\partial \eta},\frac{\partial x}{\partial \zeta} \\
\frac{\partial y}{\partial \xi},\frac{\partial y}{\partial \eta},\frac{\partial y}{\partial \zeta} \\
\frac{\partial z}{\partial \xi},\frac{\partial z}{\partial \eta},\frac{\partial z}{\partial \zeta}
\end{bmatrix}=\mathbf{X}\cdot\nabla\mathbf{N},
\end{align}
where \(
\nabla\mathbf{N} = 
\begin{bmatrix}
\frac{\partial\mathbf{N}}{\partial\xi},\frac{\partial\mathbf{N}}{\partial\eta},\frac{\partial\mathbf{N}}{\partial\zeta}
\end{bmatrix}.
\)

While a bijective mapping requires a positive Jacobian \(\lvert\mathbf{J}\rvert > 0\) everywhere within an element, it remains an open problem to check the global minimum in the parameter space. An alternative approach, popularized by the visualization tool ParaView \cite{ayachit2015paraview} and literature \cite{bracci2019hexalab, gao2019feature, guo2020cut}, is to evaluate the Jacobian at nine points: eight corners and the center (\((0.5, 0.5, 0.5)\) in the parametric space). The minimum value from these nine points is then used as a metric for element quality. The objective of the following analysis is to provide a validation check for positivity at these nine points.

Consider the scenario where the global coordinates \(\mathbf{X}\) are not fixed but are polynomial functions of some design parameters, denoted as a vector \(\mathbf{x}=(x_1,\dots,x_m)\in(0,1)^m\). Substituting such parameterized \(\mathbf{X}\) into Equation \eqref{equ:jacobianDefinition} makes the Jacobian matrix a polynomial in the \(m\) variables \(x_1,\dots,x_m\). To prove that the Jacobian is positive within domain \((0,1)\), it is needed to rewrite it as a conic combination (i.e., a linear combination with non-negative coefficients) of basis polynomials that are all non-negative in the domain. Lagrange polynomials are not such a basis. This is evident from the one-dimensional case: the only Lagrange polynomials \(\mathbf{N}(\xi)\) guaranteed to be positive over the entire interval \(\xi \in (0,1)\) are four functions \(1\), \(\xi\), \(1-\xi\), and \(\xi(1-\xi)\). Any higher-order Lagrange polynomial will become negative on some part of the interval, regardless of node placement. This limitation necessitates the use of a positive polynomial basis, such as the Bernstein basis, which is positive on \((0,1)\) for any degree. For a parameter vector \(\mathbf{x}\) and its multi-index \(\mathbf{n}=(n_1,\dots,n_m)\), where \(n_j\) denotes the highest degree for the \(j\)-th variable, the Bernstein basis is \(B_{\mathbf{i},\mathbf{n}}(\mathbf{x})=\prod_{j=1}^{m}\binom{n_j}{i_j}x_j^{i_j}(1-x_j)^{n_j-i_j}\), where \(\mathbf{x}=(x_1,\dots,x_m)\in(0,1)^m\), and \(\mathbf{i}=(i_1,\dots,i_m)\) satisfies \(0\leq i_j\leq n_j\) for all \(j\). The transformation from monomial to Bernstein coefficients is given by Theorem \ref{thm:monomialCoefficientsToBernsteinCoefficients}. The proof is given in Appendix \ref{apd:monomialCoefficientsToBernsteinCoefficients}.

\begin{theorem}
\label{thm:monomialCoefficientsToBernsteinCoefficients}
Given an integer \(m\in\mathbb{N}_{\geq1}\) and a multi-index \(\mathbf{n}=(n_1,\dots,n_m)\in\mathbb{N}^m\), for a polynomial \(\mathbf{J}(\mathbf{x})=\sum_{\mathbf{0}\leq\mathbf{i}\leq\mathbf{n}}J_{\mathbf{i}}\prod_{j=1}^{m}x_j^{i_j}\) where the parameter vector \(\mathbf{x}=(x_1,\dots,x_m)\in(0,1)^m\), the multi-index \(\mathbf{i}=(i_1,\dots,i_m)\) satisfies \(\forall j,0\leq i_j\leq n_j\). Then the Bernstein expansion \(\mathbf{J}(\mathbf{x})=\sum_{\mathbf{0}\leq\mathbf{i}\leq\mathbf{n}}\beta_{\mathbf{i}}B_{\mathbf{i},\mathbf{n}}(\mathbf{x})\), where \(B_{\mathbf{i},\mathbf{n}}(\mathbf{x})=\prod_{j=1}^{m}\binom{n_j}{i_j}x_j^{i_j}(1-x_j)^{n_j-i_j}\), has coefficients:
\begin{align}
\label{equ:bernsteinCoefficient}
\beta_\mathbf{i}=\sum_{\mathbf{0}\leq\mathbf{k}\leq\mathbf{i}}\prod_{j=1}^m\frac{\binom{i_j}{k_j}}{\binom{n_j}{k_j}}J_{\mathbf{k}},\mathbf{0}\leq\mathbf{i}\leq\mathbf{n}.
\end{align}
\end{theorem}

For a hex element whose eight vertices are given as polynomials in the parameters, Equation \eqref{equ:bernsteinCoefficient} verifies that for all multi-indices \(\mathbf{0}\leq\mathbf{i}\leq\mathbf{n}\) the coefficients \(\beta_\mathbf{i}\) are nonnegative.

\begin{table}
\centering
\caption{Coordinates of twelve red points in Figure \ref{fig:jacobian} in the local parametric space and in the global physical space.}
\vspace{-4mm}
\label{tab:jacobianCoordinates}
\begin{tabular}{c|cc}
number&local position&global position \\
\hline
\(1\)&\((s_1,0,0)\)&\(s_1(x_2,y_2,z_2)+(1-s_1)(x_1,y_1,z_1)\)\\
\(2\)&\((0,s_2,0)\)&\(s_2(x_4,y_4,z_4)+(1-s_2)(x_1,y_1,z_1)\)\\
\(3\)&\((0,0,s_3)\)&\(s_3(x_5,y_5,z_5)+(1-s_3)(x_1,y_1,z_1)\)\\
\(4\)&\((1,s_4,0)\)&\(s_4(x_3,y_3,z_3)+(1-s_4)(x_2,y_2,z_2)\)\\
\(5\)&\((1,0,s_5)\)&\(s_5(x_6,y_6,z_6)+(1-s_5)(x_2,y_2,z_2)\)\\
\(6\)&\((s_6,1,0)\)&\(s_6(x_3,y_3,z_3)+(1-s_6)(x_4,y_4,z_4)\)\\
\(7\)&\((1,1,s_7)\)&\(s_7(x_7,y_7,z_7)+(1-s_7)(x_3,y_3,z_3)\)\\
\(8\)&\((0,1,s_8)\)&\(s_8(x_8,y_8,z_8)+(1-s_8)(x_4,y_4,z_4)\)\\
\(9\)&\((s_9,0,1)\)&\(s_9(x_6,y_6,z_6)+(1-s_9)(x_5,y_5,z_5)\)\\
\(10\)&\((0,s_{10},1)\)&\(s_{10}(x_8,y_8,z_8)+(1-s_{10})(x_5,y_5,z_5)\)\\
\(11\)&\((1,s_{11},1)\)&\(s_{11}(x_7,y_7,z_7)+(1-s_{11})(x_6,y_6,z_6)\)\\
\(12\)&\((s_{12},1,1)\)&\(s_{12}(x_7,y_7,z_7)+(1-s_{12})(x_8,y_8,z_8)\)
\end{tabular}
\end{table}

\begin{figure*}
\centering
\includegraphics[width=\linewidth]{14Cases.jpg}
\vspace{-8mm}
\Description{14Cases}
\caption{14 MC base topological cases in a unit cube. The cell is tessellated by the blue surface into green 3-regular polyhedra that are kept, and gray 3-regular polyhedra that are removed. Red points are intersection points on edges. Midpoint subdividing these 3-regular polyhedra yields all-hex meshes.}
\label{fig:14Cases}
\end{figure*}

\begin{figure*}
\centering
\includegraphics[width=\linewidth]{backgroundHexTypes.jpg}
\vspace{-8mm}
\Description{backgroundHexTypes}
\caption{200 distinct background hex types in \cite{tong2026element}. No two hexes in this set can be transformed into each other by translation, rotation, or scaling.}
\label{fig:backgroundHexTypes}
\end{figure*}

\begin{figure*}
\centering
\includegraphics[width=\linewidth]{17Cases.jpg}
\vspace{-8mm}
\Description{17Cases}
\caption{17 MC complex topological cases in a unit cube can be converted to multiple base cases via local refinement. For each case: the left picture shows the blue input geometry; the middle picture shows the fixed-position local refinement template (the first compromise discussed in the text); the right picture shows the reconstructed blue surface after local refinement. Seven cases require two levels of local refinement. To simultaneously guarantee correct topology and approximation accuracy, MCHex only adopts the local refinement templates for the four cases with a * sign.}
\label{fig:17Cases}
\end{figure*}

\begin{figure}
\centering
\includegraphics[width=\linewidth]{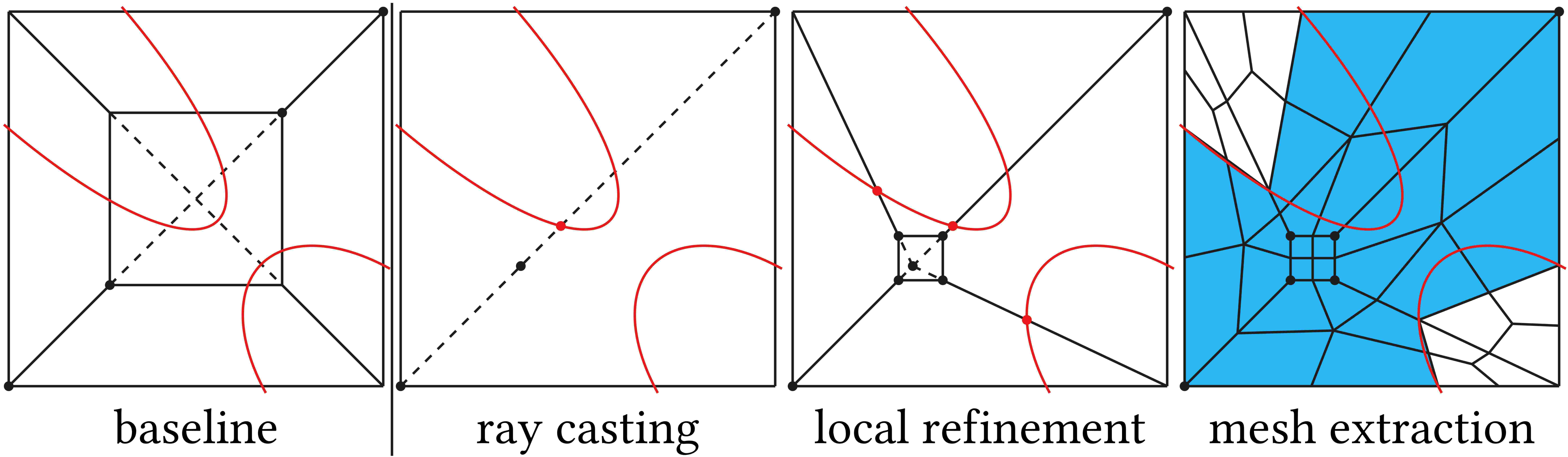}
\vspace{-8mm}
\Description{correctDDFValues}
\caption{Placement of the local refinement cell explained in two dimensions. The baseline uses fixed center and scaling, leaving ambiguity at the central sub hex. MCHex resolves this by applying ray casting to the four *-ed cases in Figure \ref{fig:17Cases}, enforcing same inside/outside status for new points and thereby ensuring that all sub background hexes fall into the base cases of Figure \ref{fig:14Cases}.}
\label{fig:correctDDFValues}
\end{figure}

To verify that the polyhedra generated by MC are all 3-regular, consider each background hex \(h\in H_{\text{Int}(T_\text{AABB})}'\) and the DDF values computed on every edge that exhibits a sign change. These DDF values lie in \([0,1]\) and represent the distance from the edge endpoint with the smaller vertex index. To prevent edges of zero length, these DDF values are clamped to \([\epsilon,1-\epsilon]\). As shown in Appendix \ref{apd:boundaryApproximationConvergenceRate}, this clamping limits the convergence rate to second order, even in flat regions of \(T_\text{geom}\) where a higher order would otherwise be achievable. In practice \(\epsilon=\frac1{1000}\) is set for most background hexes. Because DDF values almost never fall into the extreme intervals (which together cover only \(\frac1{500}\) of the edge), only a few cells for which it is difficult to guarantee a positive Jacobian require a larger \(\epsilon\). Consequently the loss is negligible.

The MC 33 classification \cite{chernyaev1995marching} enumerates all topologies that an isosurface of a trilinear interpolant can assume inside a cube, giving 33 cases. However, this classification contains a duplication: case 12.2 and case 12.3 are identical. Moreover, under mirror symmetry, cases 11 and 14 are topologically the same, leaving 31 non-repeating cases to analyze. Relaxing the surface from a trilinear isosurface to an arbitrary smooth surface introduces six additional topological cases \cite{chen2021neural}. As these do not fundamentally alter the core framework, this paper adopts the original 31 cases. Among these, the 14 base cases shown in Figure \ref{fig:14Cases} can be verified to partition the cell into several 3-regular polyhedra, each of which is hex-meshable via midpoint subdivision.

Note that Figure \ref{fig:14Cases} only depicts the situation where the background hex is a unit cube. For general background hexes in \cite{tong2026element}, the global positions of the twelve red intersection points are linear combinations of parameters \(s_i\in(0,1)\), \(i=1,\dots,12\), and the eight hex vertices \((x_i,y_i,z_i)\), \(i=1,\dots,8\) in Equation \eqref{equ:XN}; see Figure \ref{fig:jacobian} and Table \ref{tab:jacobianCoordinates}. When performing midpoint subdivision, all edge centers, face centers, and cell centers are obtained by averaging the global coordinates of the vertices belonging to that edge, face, or cell. This global averaging is essential: because the mapping from parametric space to physical space is generally not affine, the parametric midpoint is generally not the physical midpoint, and using parametric midpoints would produce a wrinkled, non-smooth physical surface.

To establish the positive Jacobian guarantee, all 200 distinct background hex configurations from \cite{tong2026element} are cataloged in Figure \ref{fig:backgroundHexTypes}; the complete test code and results are provided in the supplementary materials. For each of the 14 base cases plus the mirror-symmetric counterpart of case 13 in Figure \ref{fig:14Cases}, the 200 background hexes are iterated over. For each background hex all 24 vertex correspondences obtained by applying the octahedral group to the canonical cube corners are iterated over. This means \((14+1)\times200\times24=72,000\) tests in total. Under each test, midpoint subdivision is performed, and the resulting hex elements are written as polynomials in the edge parameters \(s_1,\dots,s_{12}\), which are initially clamped to \([\epsilon,1-\epsilon]\) with \(\epsilon=\frac1{1000}\). Using Equation \eqref{equ:bernsteinCoefficient} it is checked whether \(\min_{\mathbf{i}}\beta_{\mathbf{i}}\ge0\) holds for all coefficients; if it does, all midpoint positions are valid without any offset. If the condition fails, let \(v_P\) be the volume center of the 3-regular polyhedron that violates the condition, and let \(\{\mathbf{X}\}\) be the set of hex elements influenced by \(v_P\). Then, \(v_P\) is optimized to maximize the sum of the negative coefficient parts, \(v_P\leftarrow\arg\max_{v_P}\sum_{\mathbf{X}}\min(0,\min_\mathbf{i}\beta_\mathbf{i}(\mathbf{X}))\). When the objective reaches zero, the configuration is valid and the updated \(v_P\) is recorded; otherwise, the clamp on every edge that exhibits a sign change is tightened to \(\epsilon=\frac1{20}\), and further increased by \(\frac1{20}\) after each unsuccessful check before repeating the optimization. In practice, all cases achieve a positive Jacobian before \(\epsilon\) reaches \(0.5\). This entire optimization and verification process is performed offline. For each of the \(72,000\) tested configurations, the valid volume center offset \(v_P\) and the minimum required clamping threshold \(\epsilon\) are stored in a lookup table. During the actual meshing, MCHex queries this precomputed lookup table to warp volume centers and clamp intersection points. Although this clamping restricts the degrees of freedom for the surface in a few background hexes, in practice very few cells ever need an increased \(\epsilon\), and as shown in Appendix \ref{apd:boundaryApproximationConvergenceRate}, as long as \(\epsilon\) remains a constant the surface convergence rate stays second-order.

For the remaining 17 complex cases, shown in the left picture of each case in Figure \ref{fig:17Cases}, the complex blue surfaces make it difficult to guarantee positive Jacobians. Consequently, from an implementation standpoint, a local refinement lookup table is introduced, as illustrated in the middle picture of each case in Figure \ref{fig:17Cases}. Newly generated vertices are assigned a virtual inside/outside status, indicated by the presence or absence of a black dot. It can be verified that after local refinement every cell reduces to one of the 14 base cases. The right picture displays the reconstructed surface, with red dots marking intersection points on sign-change edges. The reconstructed surface preserves the same topology as the input geometry.

A question arises: because the inside/outside status of the newly inserted points is virtual, it may differ from the true status at those locations. Forcing these points to move so that they match the true status distorts the background hexes and makes a positive Jacobian guarantee nearly impossible. Two compromises exist. The first uses a fixed-position local refinement template; this sacrifices some surface approximation accuracy but preserves correct topology and handles all 17 complex cases. The second, adopted by MCHex, addresses face topology ambiguities using only the four cases case 4-1, case 5-1, case 7-3, and case 8-2 *-ed in Figure \ref{fig:17Cases}. The advantage is that the eight interior points of these four cases can all be assigned an inside status to obtain the correct topology. Consequently, the method illustrated in Figure \ref{fig:correctDDFValues} can simultaneously guarantee correct topology and approximation accuracy. However, the remaining 13 complex cases without a * sign are directly interpreted as base cases, which introduces incorrect topology; in practice, these cases occur very rarely, so the impact is small.

The procedure is as follows. In the baseline fixed-position refinement in the left picture of Figure \ref{fig:correctDDFValues}, the top-left and bottom-right corners lie outside the input geometry, contradicting the template that requires all interior vertices to be inside. From a known interior corner (e.g., the bottom-left one), a ray is cast toward the diagonally opposite corner; because the cell is a convex polyhedron, the entire segment lies inside the cell. The first intersection with the input geometry is found at the red point, and the local refinement center is placed one quarter of the way back toward the starting corner point. If no intersection exists, the midpoint of the segment is used. Rays are then shot from this center to every corner of the cell. For each ray the ratio of the distance to the first intersection over the total segment length is recorded; the smallest ratio among all rays is multiplied by \(\frac34\) to keep the center sub-hex safely away from the input geometry while guaranteeing that all its vertices are inside. MC followed by midpoint subdivision is then applied inside each sub background hex to extract the hex mesh.

Some further positive Jacobian proofs are necessary. Since the local refinement center can lie anywhere inside the background hex and the contraction factor can be any value in \((0,\frac34]\), proving the condition for all 200 background hex configurations is difficult in practice. Because the full set of 200 background hexes occurs very rarely, MCHex simplifies the problem by refining until only unit cubes remain as the background hex for these four types of local refinement; the corresponding proof is provided in the supplementary materials.

\section{Results and Applications}
\label{sec:resultsandApplications}

\begin{figure*}
\centering
\includegraphics[width=\linewidth]{robustness.jpg}
\vspace{-8mm}
\Description{robustness}
\caption{MCHex applied to the most challenging tire model from the Hex-Me dataset \cite{beaufort2022hex}. The top row shows the full view; the bottom row shows the zoomed-in view of the green region highlighted in the top row. From left to right: (1) input geometry \(T_\text{geom}\), (2) adaptive grid \(H_{\text{Int}(T_\text{AABB})}\) obtained by iteratively refining all cells until their misclassified volume fall below a user-provided threshold \(\epsilon_\text{vol}\), (3) conforming hex mesh \(H'_{\text{Int}(T_\text{AABB})}\) after removal of hanging nodes \cite{tong2026element}, (4) hex mesh produced by MC and midpoint subdivision for the interior \(H_{\text{Int}(T_\text{geom})}\) and the exterior \(H_{\text{Int}(T_\text{AABB})\setminus\text{Int}(T_\text{geom})}\) (\(834,802\) cells, min SJ \(8.98\times10^{-7}\), and HR \(0.798\); sharp features are smoothed out due to the limitation of MC), and (5) interior mesh \(H_{\text{Int}(T_\text{geom})}\) after post-processing smoothing \cite{protais2026versatile}, with min SJ \(0.0191\), and HR \(0.354\), where sharp features are preserved.}
\label{fig:robustness}
\end{figure*}

To validate the MCHex approach, a C++ prototype is implemented following Algorithm \ref{alg:entirePipeline}. Figure \ref{fig:robustness} demonstrates the application of MCHex to the most challenging tire model from the Hex-Me dataset \cite{beaufort2022hex}, a shape for which no prior hex meshing method has successfully produced a mesh. Beyond this demonstration of robustness, the MCHex-raw output and the output after smoothing with \cite{protais2026versatile} are evaluated against six previous state-of-the-art methods \cite{marechal2016all,gao2019feature,dumery2022evocube,liu2023locally,mestrallet2025validity,protais2025automatic} on a benchmark dataset of 202 geometries \cite{gao2019feature}. The number of hex mesh vertices (\#vert) and \#hex are reported. Boundary fidelity is quantified by the HR and IoU. HR captures the surface deviation between the input surface \(T_\text{geom}\) and the mesh boundary \(\partial H_{\text{Int}(T_\text{geom})}\); IoU measures the volumetric overlap between the interior of the input geometry \(\text{Int}(T_\text{geom})\) and the hex mesh. The runtime is reported. Mesh quality is assessed using five metrics: minimum scaled Jacobian (min SJ), minimum shape (min SP), maximum aspect Frobenius (max AF), maximum skew (max SK), and minimum shear (min SH). Experimental results show that, with the same or fewer elements, MCHex-raw produces meshes faster but with slightly worse boundary error and mesh quality than previous methods. After smoothing, MCHex achieves similar boundary error and mesh quality to the prior work. It is further shown that the runtime of MCHex-raw is linear in \#hex and sublinear in \#tri, and peak RAM is linear in both \#hex and \#tri. Finally, it is demonstrated that the raw meshes generated by MCHex can be post-processed via singularity simplification to reduce mesh singularities and are suitable for finite element simulations.

\begin{figure}
\centering
\includegraphics[width=\linewidth]{compareMCHexRawToOthers.jpg}
\vspace{-8mm}
\Description{compareMCHexRawToOthers}
\caption{Comparison of MCHex-raw (x-axis) with six prior methods (y-axes), showing the first 10 metrics in Table \ref{tab:compareMCHexToOthers}. Each point represents a test model. MCHex-raw outperforms if points fall inside the shaded region bounded by the dashed \(y=x\) line. The diagonal may not span corners due to different axis scales, and curves are shown in the last 5 rows because the x-axis is logarithmic (vs. linear y-axis) to evenly distribute data.}
\label{fig:compareMCHexRawToOthers}
\end{figure}

\begin{figure}
\centering
\includegraphics[width=\linewidth]{compareMCHexSmoothedToOthers.jpg}
\vspace{-8mm}
\Description{compareMCHexSmoothedToOthers}
\caption{Comparison of MCHex-smoothed (x-axis) with six prior methods (y-axes). The rows for \#vert and \#hex are identical to those in Figure \ref{fig:compareMCHexRawToOthers} and are therefore omitted. The remaining convention is the same as in Figure \ref{fig:compareMCHexRawToOthers}.}
\label{fig:compareMCHexSmoothedToOthers}
\end{figure}

\setlength{\dashlinedash}{1.35pt}
\setlength{\dashlinegap}{1.35pt}
\begin{table*}
\centering
\caption{Comparison of the eight methods over metrics. For the first 10 metrics, each cell reports the median ratio of a method's value relative to MCHex-raw (left) and MCHex-smoothed (right) across all available input geometries. Bold text indicates the best ratio. Arrows \(\uparrow\)/\(\downarrow\) indicate whether higher or lower values are desirable. Min SJ in the range \([-1,1]\) is increased by 1 to make ratios meaningful; max AF in the range \([1,\infty)\) is replaced by its reciprocal to keep ratios bounded. The last two rows report the meshing and simulation success rates: the fraction of the 202 input geometries for which a method produces a valid hex mesh, and the fraction of those meshes that achieve convergence in simulation.}
\vspace{-4mm}
\label{tab:compareMCHexToOthers}
\begin{tabular}{c|cccccccc}
metrics&AlgoHex&Evocube&Gao et al.&Hexotic&marchinghex&\makecell{Validity-\\first}&\makecell{MCHex-\\raw}&\makecell{MCHex-\\smoothed}\\
\hline
\#vert\(\downarrow\)&\(1.08/1.08\)&\(1.18/1.18\)&\(1.06/1.06\)&\(2.58/2.58\)&\(1.14/1.14\)&\(1.19/1.19\)&\(\mathbf{1.00}/\mathbf{1.00}\)&\(\mathbf{1.00}/\mathbf{1.00}\)\\
\#hex\(\downarrow\)&\(1.08/1.08\)&\(1.16/1.16\)&\(1.05/1.05\)&\(2.67/2.67\)&\(1.15/1.15\)&\(1.17/1.17\)&\(\mathbf{1.00}/\mathbf{1.00}\)&\(\mathbf{1.00}/\mathbf{1.00}\)\\
HR\(\downarrow\)&\(0.613/4.98\)&\(0.631/3.14\)&\(0.245/1.12\)&\(0.295/1.42\)&\(0.898/4.92\)&\(0.480/2.44\)&\(1.00/4.74\)&\(\mathbf{0.211}/\mathbf{1.00}\)\\
IoU\(\uparrow\)&\(1.04/0.984\)&\(1.03/0.995\)&\(1.039/0.998\)&\(1.03/0.995\)&\(1.02/0.983\)&\(1.03/0.996\)&\(1.00/0.960\)&\(\mathbf{1.041}/\mathbf{1.00}\)\\
runtime\(\downarrow\)&\(5,330/187\)&\(20.4/0.478\)&\(1,300/35.6\)&\(1.71/0.0469\)&\(35.4/0.728\)&\(28.5/0.679\)&\(\mathbf{1.00}/\mathbf{0.0246}\)&\(40.6/1.00\)\\
min SJ\(+1\)\(\uparrow\)&\(0.0106/0.0142\)&\(0.597/0.574\)&\(1.15/\mathbf{1.03}\)&\(1.07/0.953\)&\(\mathbf{1.16}/1.02\)&\(0.828/0.708\)&\(1.00/0.886\)&\(1.13/1.00\)\\
min SP\(\uparrow\)&\(0/0\)&\(0/0\)&\(\mathbf{1,020}/\mathbf{1.06}\)&\(592/0.637\)&\(810/0.966\)&\(0/0\)&\(1.00/0.00111\)&\(901/1.00\)\\
\(\frac1{\text{max AF}}\)\(\uparrow\)&\(0/0\)&\(0/0\)&\(\mathbf{321}/\mathbf{1.10}\)&\(103/0.391\)&\(201/0.852\)&\(0/0\)&\(1.00/0.00391\)&\(256/1.00\)\\
max SK\(\downarrow\)&\(0.815/0.931\)&\(0.962/1.08\)&\(\mathbf{0.798}/\mathbf{0.914}\)&\(0.835/0.948\)&\(0.929/1.05\)&\(0.939/1.05\)&\(1.00/1.14\)&\(0.875/1.00\)\\
min SH\(\uparrow\)&\(0/0\)&\(0/0\)&\(\mathbf{3,530}/\mathbf{1.25}\)&\(925/0.508\)&\(2,330/1.19\)&\(0/0\)&\(1.00/0.000437\)&\(2,290/1.00\)\\
\arrayrulecolor[HTML]{AAAAAA}
\hdashline
\makecell{meshing\\success rate}\(\uparrow\)&\(0.742\)&\(0.995\)&\(\mathbf{1.00}\)&\(\mathbf{1.00}\)&\(\mathbf{1.00}\)&\(0.985\)&\(\mathbf{1.00}\)&\(\mathbf{1.00}\)\\
\hdashline
\arrayrulecolor{black}
\makecell{simulation\\success rate}\(\uparrow\)&\(0.581\)&\(0.772\)&\(0.946\)&\(\mathbf{0.975}\)&\(0.926\)&\(0.782\)&\(0.946\)&\(0.950\)\\
\end{tabular}
\end{table*}

The comparison between MCHex-raw and the six prior methods is visualized in Figure \ref{fig:compareMCHexRawToOthers}. The first two rows compare \#vert and \#hex. Usually, a higher vertex/element count yields better geometric fidelity but also increases downstream simulation cost; therefore each method's budget is controlled as follows. Because Hexotic \cite{marechal2016all} cannot be obtained due to exclusive rights, and \cite{gao2019feature} exhausts the 64GB total memory, distorting the latter's runtime measurement, the publicly available results of the two methods from \cite{gao2019feature} are directly used, and their runtimes are scaled based on single-threaded CPU throughput. For MCHex, the \(\epsilon_\text{vol}\) threshold is iteratively adjusted until \#hex is slightly below the smaller of the element counts produced by \cite{gao2019feature} and \cite{marechal2016all}. The other four methods all require a tet mesh, which is generated with \cite{hu2020fast} using ``-l 0.02'' to slightly refine the tet mesh because coarser meshes cause \cite{liu2023locally} to fail, while further refinement makes it prohibitively slow. After tet meshes are generated, refinement parameters of the four hex meshing methods are tuned so that their overall element counts slightly exceed those of MCHex. All other settings are kept at their default values. As shown in the first two rows of Table \ref{tab:compareMCHexToOthers}, all ratios are above 1, which means this budget control is successful.

Table \ref{tab:compareMCHexToOthers} and the fifth row of Figure \ref{fig:compareMCHexRawToOthers} show that MCHex-raw processes inputs faster than all competing methods. The sixth row of the figure confirms that its min SJ is always positive, and the eleventh row of the table demonstrates that it is guaranteed to produce a hex mesh on every input. This is because MC's Occam's razor principle only requires a reasonable DDF to produce a mesh. In the last row of the table, the simulation success rate of MCHex-raw is comparable to that of the optimized meshes. Nevertheless, MCHex has flaws. On the two boundary-approximation metrics HR and IoU, it is defeated by most other methods, because its surface is essentially the unprojected MC surface, which inherently loses sharp features. On the five mesh-quality metrics, MCHex-raw performs worse than some other heuristic grid-based methods: its positive Jacobian guarantee does nothing to promote other quality measures.

Applying the smoothing method \cite{protais2026versatile} to MCHex-raw results in MCHex-smoothed. This yields substantial gains in boundary approximation and mesh quality. As shown in Figure \ref{fig:compareMCHexSmoothedToOthers} and Table \ref{tab:compareMCHexToOthers}, MCHex-smoothed ranks first among all methods on HR and IoU, third on min SJ, max SK, and min SH, and second on min SP and max AF. This demonstrates that third-party post-processing can lift MCHex-raw meshes to a quality level comparable with prior methods. However, smoothing takes significantly longer: it places sixth overall in runtime.

\begin{figure*}
\centering
\includegraphics[width=\linewidth]{selectedComparisons.jpg}
\vspace{-8mm}
\Description{selectedComparisons}
\caption{A random subset of models from the benchmark \cite{gao2019feature}. Input geometry and eight outputs (left: blue surface, right: SJ colormap) are shown.}
\label{fig:selectedComparisons}
\end{figure*}

Figure \ref{fig:selectedComparisons} presents a comparison on a random subset of 10 geometries out of the 202 input models. The leftmost column displays the input geometry, while the remaining columns display the hex meshes generated by eight evaluated methods. For each mesh, the left half displays the surface, and the right half displays the SJ colormap. The frame-field method, AlgoHex \cite{liu2023locally}, typically yields the fewest singularities and a uniform refinement. However, it is the least robust, being highly prone to failure, poor boundary approximation, and negative Jacobians. Similarly, the polycube-based methods, Evocube \cite{dumery2022evocube} and Validity-first \cite{mestrallet2025validity}, generate the second fewest singularities with uniform refinement, but are the second least robust, occasionally suffering from poor boundary approximation and negative Jacobians. The remaining grid-based methods produce significantly more singularities. Their internal refinement is either uniform, as in marchinghex \cite{protais2025automatic}, or adaptive, as in \cite{gao2019feature}, Hexotic \cite{marechal2016all}, and MCHex. These approaches are the most robust and rarely fail. However, for simple geometries where frame-field or polycube methods succeed, the grid-based methods often yield a lower min SJ.

\begin{figure}
\centering
\includegraphics[width=\linewidth]{timeSpaceComplexity.jpg}
\vspace{-8mm}
\Description{timeSpaceComplexity}
\caption{Runtime and peak RAM of MCHex-raw on the 202 input geometries from \cite{gao2019feature} and the 189 input geometries from \cite{beaufort2022hex}. Scatter plots show the quantities as functions of \#hex and \#tri. To augment the data, for each input model a series of hex meshes is generated with misclassified volume thresholds \(\epsilon_\text{vol}=10^{0},10^{-1},\dots,10^{-8}\). When \#hex is large, both runtime and peak RAM scale linearly with it. As a function of \#tri, runtime grows sublinearly, while peak RAM grows linearly.}
\label{fig:timeSpaceComplexity}
\end{figure}

The runtime and peak RAM of MCHex-raw are analyzed as functions of \#hex and \#tri in Figure \ref{fig:timeSpaceComplexity}. To obtain a wide distribution of data points, 391 input geometries are used: 202 from \cite{gao2019feature} and 189 from \cite{beaufort2022hex}. For each model, a series of hex meshes is generated with misclassified volume thresholds \(\epsilon_\text{vol}=10^{0},10^{-1},\dots,10^{-8}\), yielding \(391\times9=3,519\) data points in total. As \#hex increases, runtime and peak RAM initially exhibit superlinear growth, but the trend transitions to roughly linear once \#hex exceeds \(10^6\). With increasing \#tri, after \(10^5\) triangles the runtime growth becomes sublinear, likely because the implementation stores triangle indices in local grid cells and ray casting can then skip large portions of the input triangles; meanwhile peak RAM growth remains reasonably linear. Direct peak RAM comparisons with prior methods are difficult because the results provided in \cite{gao2019feature} for its own method and for \cite{marechal2016all} do not include peak RAM usage. Nevertheless, in local tests the method of \cite{gao2019feature} frequently exhausts 64GB of RAM when generating hex meshes on the order of tens to hundreds of thousands of elements, while MCHex-raw uses only about 10GB even for meshes with tens of millions of hex elements. Overall, MCHex-raw demonstrates strong runtime and RAM efficiency, and because no dedicated performance optimizations are applied, substantial room for further improvement remains.

\begin{figure}
\centering
\includegraphics[width=\linewidth]{simplification.jpg}
\vspace{-8mm}
\Description{simplification}
\caption{Singularity simplification on 10 random models from the benchmark \cite{gao2019feature} using the method of \cite{gao2017robust}. For each model, the left and right halves show the mesh before and after simplification, respectively. Singular edges are highlighted in red.}
\label{fig:simplification}
\end{figure}

The hex meshes generated by MCHex can be further improved through post-processing singularity structure simplification \cite{gao2017robust}. Figure \ref{fig:simplification} illustrates the singularity structures of 10 randomly selected models before and after simplification. Following the definition in \cite{pietroni2022hex}, red edges denote singular edges. Interior edges are shared by a number of hex elements not equal to four, and surface edges are shared by a number not equal to two. The singularity complexity of MCHex-generated meshes can be effectively reduced using \cite{gao2017robust}. This provides an alternative approach to generating semi-structured meshes beyond frame-field and polycube methods, although the current number of singular edges remains relatively high, especially on non-axis-aligned geometries.

\begin{figure*}
\centering
\includegraphics[width=\linewidth]{simulation.jpg}
\vspace{-8mm}
\Description{simulation}
\caption{A random subset of models from the benchmark \cite{gao2019feature}. For each model, the leftmost picture is the input geometry, and the remaining eight columns show the FEA results for solving a nonlinear Poisson equation on hex meshes generated by different methods. Models shown in blue indicate that the simulation fails to converge.}
\label{fig:simulation}
\end{figure*}

To compare the generated meshes in downstream applications, FEA is conducted on all produced hex meshes. Specifically, a second-order nonlinear Poisson equation \(\Delta u=u^2\) with \(u=1\) on \(\partial\Omega\) is linearized and solved iteratively using the Newton-Raphson method, where \(\Omega\) denotes the three-dimensional volumetric domain of the input geometry. This specific equation is chosen as it is fundamentally challenging for Monte Carlo-based PDE solving methods that do not rely on meshes \cite{muller1956some}. It should be noted that the FEA solver used here is not the most robust one; however, since the same solver is applied to all methods, the relative comparison remains meaningful. The last row of Table \ref{tab:compareMCHexToOthers} reports the simulation success rate (a.k.a. the FEA solver convergence rate) for all evaluated methods. Figure \ref{fig:simulation} shows the results for 10 randomly selected models. In this figure, the two failures recorded for AlgoHex result from failing to produce an output mesh, rather than simulation divergence. Grid-based approaches generally achieve the highest success rates. Notably, MCHex-raw also exhibits a high convergence rate comparable to that of all other smoothed methods, suggesting that MCHex-raw meshes are already suitable for certain finite element simulations. However, it is important to emphasize that these simulation outcomes alone cannot conclusively determine the overall superiority of a method's mesh quality. This is because different methods yield varying boundary approximation errors. Surface fidelity and mesh quality are a trade-off. Furthermore, differences in mesh density across methods may affect the simulation accuracy. Finally, these findings are specific to the tested nonlinear Poisson equation and may not universally generalize to all PDE problems.

\section{Conclusion and Future Work}
\label{sec:conclusionandFutureWork}

This paper introduces MCHex, a new paradigm for boundary approximation in grid-based hex meshing. It replaces the conventional pipeline of outside element removal, padding, and heuristic projection with robust MC cutting and midpoint subdivision. Founded on MC, 3-regular graphs, midpoint subdivision, and Bernstein coefficient checks, MCHex automatically pads the boundary while guaranteeing all-positive Jacobians, second-order boundary convergence, and a manifold boundary for arbitrarily complex geometries. Extensive experiments demonstrate that with the same or fewer elements, MCHex-raw generates meshes much faster than previous art, albeit with slightly worse boundary error and mesh quality. However, after smoothing, MCHex achieves the lowest boundary error and mesh quality comparable to existing methods. Furthermore, the algorithm demonstrates favorable time and space complexity. Finally, the singularity structure of MCHex meshes can be simplified, and the MCHex meshes are suitable for finite element simulations.

\begin{figure}
\centering
\includegraphics[width=\linewidth]{awayAreaDontRefine.jpg}
\vspace{-8mm}
\Description{awayAreaDontRefine}
\caption{Future approach for saving elements, illustrated in two dimensions and applicable only to 2-refinement. Top row (left to right): (1) the red input geometry; (2) adaptive grid refinement satisfying the moderate balance and pairing rules. Middle row: (3) current MCHex result with hanging nodes removed; (4) global propagation of midpoint subdivision across all grid cells. Bottom row: (5) the improved method first virtually subdivides all grid cells intersected by the input geometry together with their sibling cells; (6) the moderate balance and pairing rules are enforced on the remaining cells and hanging nodes are removed; (7) MCHex midpoint subdivision is applied only to the cells marked by the virtual subdivision scaffolds, after which the scaffolds are removed. This prevents unnecessary midpoint subdivision from propagating throughout the entire grid.}
\label{fig:awayAreaDontRefine}
\end{figure}

While effective, MCHex presents certain limitations that offer promising directions for future work. First, the midpoint subdivision strategy inherently multiplies the local hex element count by approximately eight with no improvement in topology. This overhead means that, under the same grid resolution, MCHex generates more elements than \cite{gao2019feature} for some models, as shown in the first row of Figure \ref{fig:compareMCHexRawToOthers}. This limitation could be mitigated by adopting an element-saving refinement approach, as illustrated in Figure \ref{fig:awayAreaDontRefine}. This approach relies on a key observation: under 2-refinement, when cells intersected by the input geometry undergo midpoint subdivision, their boundary quads are always split into four smaller squares. This can be seen as a virtual subdivision of the intersected cells. Therefore, a virtual subdivision of these intersected cells can be performed. Due to the pairing rule, their sibling cells are automatically subdivided as well. Then the moderate balance and pairing rules are applied to the rest of the grid, and hanging nodes are removed using a potential 2-refinement version of \cite{tong2026element}. Finally, applying MC cutting and midpoint subdivision to the intersected cells yields boundaries that naturally conform to the surrounding cells. Compared to the global splitting in MCHex, this prevents the midpoint subdivision from propagating unnecessarily throughout the entire grid.

Second, although sharp features can be preserved during the smoothing phase using \cite{protais2026versatile}, MCHex-raw is fundamentally limited by the underlying MC algorithm and cannot preserve sharp features. Developing a feature-preserving variant of the cutting step remains a direction for future work; specifically, integrating machine learning approaches such as Neural Marching Cubes \cite{chen2021neural} or Neural Dual Contouring \cite{chen2022neural} can effectively reconstruct sharp features.

Third, the current Bernstein coefficient check provides a sufficient but not necessary condition for positive Jacobians. Even when negative coefficients exist, the Jacobian may still be positive across the entire domain. Future work could explore tighter proofs to reduce unnecessary subdivisions and further improve efficiency.

\begin{acks}
%H. Tong and Y. J. Zhang were supported in part by the NSF grants (CMMI-1953323 and CBET-2332084).
\end{acks}

\bibliographystyle{ACM-Reference-Format}
\bibliography{bibliography.bib}

\appendix
\section{Boundary Approximation Convergence Rate}
\label{apd:boundaryApproximationConvergenceRate}

Some of the notations follow definitions in Section \ref{sec:algorithmOverview}.

\begin{theorem}
\label{thm:boundaryApproximationConvergenceRate}

Let \(S_\text{geom}=\{x\in\mathbb{R}^3\mid\phi(x)=0\}\) be a closed \(C^2\) compact surface, where \(\phi:\mathbb{R}^3\to\mathbb{R}\) is a signed distance function. Therefore it holds that \(\lVert\nabla\phi(x)\rVert_2=1\). Let \(H\) denote a uniform cubic grid of size \(h\) so that \(H\) contains \(S_\text{geom}\). \(h_{i,j,k}\) denotes the grid cell occupying the volume \((ih,jh,kh)-(ih+h,jh+h,kh+h)\), where \(i,j,k\in\mathbb{N}\). Let \(T_\text{baseline}\), \(T_\text{midpoint}\), and \(T_\text{MCHex}\) denote the reconstructed piecewise linear triangle meshes using the baseline removing outside element method (e.g., \cite{marechal2016all,gao2019feature}), midpoint MC \cite{protais2025automatic}, and MCHex on \(H\). The baseline method keeps a cell if and only if more than four of its corners are inside \(S_\text{geom}\). The midpoint MC method uses MC, whereas all intersection points are at the middle of the edges. MCHex extracts MC surfaces with exact intersection points on the grid edges. Under this setting, the HD \(d(S_\text{geom},T)\) between \(S_\text{geom}\) and the approximated surfaces satisfies:
\begin{align}
d(S_\text{geom},T_\text{baseline})&\in\Theta(h),\label{equ:baselineConvergenceRate}\\
d(S_\text{geom},T_\text{midpoint})&\in\Theta(h),\label{equ:midpointConvergenceRate}\\
d(S_\text{geom},T_\text{MCHex})&\in\Theta(h^2),\label{equ:MCHexConvergenceRate}
\end{align}
where \(\Theta(h)\) and \(\Theta(h^2)\) denote the sets of functions exhibiting exact linear and quadratic convergence rates, respectively \cite{knuth1976big}.
\end{theorem}

\begin{proof}

HD is defined as
\begin{align}
\label{equ:hausdorffDistance}
d(A,B)=\max(\sup_{x\in A}\inf_{y\in B}\lVert x-y\rVert_2,\sup_{y\in B}\inf_{x\in A}\lVert x-y\rVert_2).
\end{align}
The property of \(\phi\) is then given as follows:
\begin{align}
\label{equ:SDF}
\lvert \phi(x)\rvert=\inf_{y\in S_\text{geom}}\lVert x-y\rVert_2.
\end{align}

\addvspace{1em}
To prove Equation \eqref{equ:baselineConvergenceRate}, it is necessary to show that both the upper and the lower bounds of \(d(S_\text{geom},T_\text{baseline})\) are linear as \(h\to0\).

\addvspace{1em}
\noindent\textbf{(1) Upper bound of \(d(S_\text{geom},T_\text{baseline})\)}

For any \(y\in T_\text{baseline}\), it always lies on the shared face of a kept grid cell and a removed grid cell. The union of these two cells always contains both inside points and outside points. From the continuity of \(\phi\), there exists a point \(x\) in this union so that \(\phi(x)=0\). Since \(x\) and \(y\) are both in the union, \(\lVert x-y\rVert_2\leq\sqrt{h^2+h^2+(2h)^2}=\sqrt6h\) holds.
From \(\phi(x)=0\) and the definition of \(\phi\), \(\lvert\phi(y)\rvert=\lvert\phi(y)-\phi(x)\rvert\leq\lVert y-x\rVert_2\leq\sqrt6h\) is obtained. Using Equation \eqref{equ:SDF} and taking the supremum of \(y\) over \(T_\text{baseline}\), \(\sup_{y\in T_\text{baseline}}\inf_{x\in S_\text{geom}}\lVert x-y\rVert_2\leq\sqrt6h\) can be obtained.

For any \(x_0\in S_\text{geom}\), let the grid cell containing it be \(h_1\). Notice that \(\max(\lvert\frac{\partial\phi(x_0)}{\partial x}\rvert,\lvert\frac{\partial\phi(x_0)}{\partial y}\rvert,\lvert\frac{\partial\phi(x_0)}{\partial z}\rvert)\geq\frac1{\sqrt3}\). Without loss of generality, it is assumed that \(\frac{\partial\phi(x_0)}{\partial z}\geq\frac1{\sqrt3}\).
Since \(S_\text{geom}\) is \(C^2\), \(\phi\) is also \(C^2\) in the tubular neighborhood of \(S_\text{geom}\), which means \(\nabla\phi\) and \(\nabla^2\phi\) are continuous in this tubular neighborhood. Therefore, \(\exists r>0\) s.t. \(\forall y\) satisfying \(\lVert y-x_0\rVert_2\leq r\), there exists a lower bound \(\frac{\partial\phi(y)}{\partial z}\geq\frac1{2\sqrt3}\).

Let \(z\) be any corner point of \(h_1\). Since \(x_0\in h_1\), the distance from \(x_0\) to \(z\) does not exceed the diagonal length of \(h_1\): \(\lVert z-x_0\rVert_2\leq\sqrt3h\). From \(\phi(x_0)=0\) and the definition of \(\phi\), it can be obtained that \(\lvert\phi(z)\rvert=\lvert\phi(z)-\phi(x_0)\rvert\leq\lVert z-x_0\rVert_2\leq\sqrt3h,-\sqrt3h\leq\phi(z)\leq\sqrt3h\).
Let \(z'=z+\Delta kh\hat z\), where \(\Delta k\in\mathbb{Z}\). By the mean value theorem, there exists \(\xi=z+\lambda\Delta kh\hat z,\lambda\in[0,1]\) s.t. \(\phi(z')=\phi(z)+\Delta kh\frac{\partial\phi(\xi)}{\partial z}.\)

When \(\Delta k=7\), assume \(h\) is sufficiently small to ensure that \(\xi\) is in the neighborhood of \(x_0\) with radius \(r\), then \(\phi(z')\geq\phi(z)+\frac{7h}{2\sqrt3}\geq-\sqrt3h+\frac{7h}{2\sqrt3}>0\). Conversely, when \(\Delta k=-7\), then \(\phi(z')\leq\phi(z)-\frac{7h}{2\sqrt3}\leq\sqrt3h-\frac{7h}{2\sqrt3}<0\). This means the seventh grid cell above \(h_1\) has eight corners outside \(S_\text{geom}\) and it is removed. The seventh grid cell below \(h_1\) has eight corners inside \(S_\text{geom}\) and it is kept. Within this discrete sequence of 15 grid cells, there must exist a pair of adjacent grid cells such that the upper one is removed and the lower one is kept. The face shared by this adjacent pair belongs to \(T_\text{baseline}\). Assuming \(y\) is now on this face, the distance from \(y\) to \(x_0\) satisfies \(\lVert y-x_0\rVert_2\leq\sqrt{(7h)^2+h^2+h^2}=\sqrt{51}h\). Then it is obtained that \(\sup_{x\in S_\text{geom}}\inf_{y\in T_\text{baseline}}\lVert x-y\rVert_2\leq\sqrt{51}h\).
From Equation \eqref{equ:hausdorffDistance}, it follows that \(d(S_\text{geom},T_\text{baseline})\leq\sqrt{51}h\) or the upper bound of \(d(S_\text{geom},T_\text{baseline})\) is linear.

\addvspace{1em}
\noindent\textbf{(2) Lower bound of \(d(S_\text{geom},T_\text{baseline})\)}

Since \(S_\text{geom}\) is a closed \(C^2\) compact surface in \(\mathbb{R}^3\), its Gauss map is surjective. Thus, \(\exists x_0\in S_\text{geom}\) such that \(\nabla\phi(x_0)=
\begin{bmatrix}
\frac1{\sqrt3},\frac1{\sqrt3},\frac1{\sqrt3}
\end{bmatrix}^\top\).
From the bound proven above, it is shown that \(\inf_{y\in T_\text{baseline}}\lVert x_0-y\rVert_2\leq\sqrt{51}h\). Let \(y_0\in T_\text{baseline}\) be a point that achieves this infimum (i.e., \(\lVert y_0-x_0\rVert_2\leq\sqrt{51}h\)), then the point \(y_0\) must lie on a face \(f\subset T_\text{baseline}\). Without loss of generality, assume \(f\) is an axis-aligned face spanning \((x_1,y_1,z_1)-(x_1+h,y_1+h,z_1)\). Let \(p\in f\) be any point on this face with coordinates \((p_x,p_y,z_1)\). Its distance to \(x_0\) is bounded by \(\lVert p-x_0\rVert_2\leq\lVert p-y_0\rVert_2+\lVert y_0-x_0\rVert_2\leq\sqrt2h+\sqrt{51}h\).

The first-order Taylor expansion of \(\phi\) at \(x_0(x_{0,x},x_{0,y},x_{0,z})\) is 
\begin{align}
\label{equ:taylorExpansion}
\phi(p)&=\phi(x_0)+\nabla\phi(x_0)\cdot(p-x_0)+\frac12(p-x_0)^\top\nabla^2\phi(\xi)(p-x_0)\nonumber\\
&=\frac1{\sqrt3}(p_x-x_{0,x}+p_y-x_{0,y}+p_z-x_{0,z})+R_2(p)\nonumber\\
&=\frac{g(p_x,p_y)}{\sqrt3}+R_2(p),
\end{align}
where \(\xi=x_0+\lambda(p-x_0),\lambda\in[0,1]\).
Continuous function \(\nabla^2\phi\) on compact domains must have bounds, therefore \(\exists M>0\) s.t. for all \(\xi\) in this tubular neighborhood, one has \(\lVert\nabla^2\phi(\xi)\rVert_2\leq M\). Therefore, it can be derived that \(\lvert R_2(p)\rvert\leq\frac12\lVert\nabla^2\phi(\xi)\rVert_2\lVert p-x_0\rVert_2^2\leq\frac12M\lVert p-x_0\rVert_2^2\leq\frac12M(\sqrt2+\sqrt{51})^2h^2\).

\(g(p_x,p_y)\) is linear, therefore its extreme values on the square \(f\) occur at two corners \((x_1,y_1,z_1)\) and \((x_1+h,y_1+h,z_1)\). The difference between these extreme values is \(g(x_1+h,y_1+h)-g(x_1,y_1)=2h\). Therefore
\begin{align}
&\max(\lvert g(x_1,y_1)\rvert,\lvert g(x_1+h,y_1+h)\rvert)\nonumber\\\geq&\frac12(\lvert g(x_1,y_1)\rvert+\lvert g(x_1+h,y_1+h)\rvert)\nonumber\\\geq&\frac12(g(x_1+h,y_1+h)-g(x_1,y_1))\nonumber\\=&h.\nonumber
\end{align}
Combining this with Equation \eqref{equ:taylorExpansion}, there exists \(p\) at one of the two corners, \((x_1,y_1,z_1)\) or \((x_1+h,y_1+h,z_1)\) such that \(\lvert\phi(p)\rvert\geq\frac{\lvert g(p_x,p_y)\rvert}{\sqrt3}-\frac12M(\sqrt2+\sqrt{51})^2h^2\geq\frac{h}{\sqrt3}-\frac12M(\sqrt2+\sqrt{51})^2h^2\). Using Equation \eqref{equ:SDF}, the shortest distance from \(p\) to \(S_\text{geom}\) satisfies \(\inf_{x\in S_\text{geom}}\lVert p-x\rVert_2=\lvert\phi(p)\rvert\). Taking the supremum over all points \(y\in T_\text{baseline}\) in Equation \eqref{equ:hausdorffDistance}, it follows that \(d(S_\text{geom},T_\text{baseline})\geq\sup_{y\in T_\text{baseline}}\inf_{x\in S_\text{geom}}\lVert x-y\rVert_2\). For a sufficiently small grid size \(h\), the linear term \(\frac h{\sqrt3}\) dominates the quadratic term \(\frac12M(\sqrt2+\sqrt{51})^2h^2\).

With both the upper and the lower bounds of \(d(S_\text{geom},T_\text{baseline})\) being linear as \(h\to0\), Equation \eqref{equ:baselineConvergenceRate} is proved.

\addvspace{1em}
Similarly to prove Equation \eqref{equ:midpointConvergenceRate}, it is necessary to show that both the upper and the lower bounds of \(d(S_\text{geom},T_\text{midpoint})\) are linear as \(h\to0\).

\addvspace{1em}
\noindent\textbf{(3) Upper bound of \(d(S_\text{geom},T_\text{midpoint})\)}

For any \(y\in T_\text{midpoint}\), it lies in a triangle generated by the midpoint MC rule inside a grid cell \(h_1\) of \(H\). Since the surface passes through \(h_1\), the cell contains at least one point \(x\) with \(\phi(x)=0\). Both \(y\) and \(x\) belong to \(h_1\). Hence \(\inf_{x\in S_\text{geom}}\lVert y-x\rVert_2\leq\sqrt3h\). Taking the supremum gives \(\sup_{y\in T_\text{midpoint}}\inf_{x\in S_\text{geom}}\lVert x-y\rVert_2\leq\sqrt3h.\)

Conversely, for any \(x\in S_\text{geom}\), let \(h_1\) be the grid cell containing \(x\). Similar to the gradient bound established in the \(T_\text{baseline}\) proof, there exists a nearby grid cell within \(7h\) distance where a sign change occurs. The midpoint MC algorithm will therefore produce at least one triangle in the z-axis interval \([-\frac{13}2h,\frac{13}2h]\). Picking \(y\) on that triangle gives \(\lVert x-y\rVert_2\leq\sqrt{(\frac{13}2h)^2+h^2+h^2}=\frac{\sqrt{177}}2h\). Thus, it is guaranteed that \(\sup_{x\in S_\text{geom}}\inf_{y\in T_\text{midpoint}}\lVert x-y\rVert_2\leq\frac{\sqrt{177}}2h\). Combining both directions, the HD in Equation \eqref{equ:hausdorffDistance} satisfies \(d(S_\text{geom},T_\text{midpoint})\leq\frac{\sqrt{177}}2h\). In other words, the upper bound of \(d(S_\text{geom},T_\text{midpoint})\) is linear.

\addvspace{1em}
\noindent\textbf{(4) Lower bound of \(d(S_\text{geom},T_\text{midpoint})\)}

Since \(S_\text{geom}\) is a closed \(C^2\) compact surface, its Gauss map is surjective. Therefore, \(\exists x\in S_\text{geom}\) such that \(\nabla\phi(x)=
\begin{bmatrix}
\frac23,\frac23,\frac13
\end{bmatrix}^\top\).
From the upper bound proven above, one has \(\inf_{y\in T_\text{midpoint}}\lVert x-y\rVert_2\leq\frac{\sqrt{177}}2h\). Let \(y_0\in T_\text{midpoint}\) be a point that achieves this infimum. \(y_0\) must lie on a face \(f\subset T_\text{midpoint}\) generated within a grid cell \(h_1\). The vertices of \(f\), denoted as \(p_1,p_2,p_3\), are midpoints of the edges of \(h_1\) where \(\phi\) changes sign at two endpoints. Any point \(p(p_x,p_y,p_z)\in f\) has its distance to \(x\) bounded by \(\lVert p-x\rVert_2\leq\lVert p-y_0\rVert_2+\lVert y_0-x\rVert_2\leq\sqrt3h+\frac{\sqrt{177}}2h=Ch\).

The first-order Taylor expansion of \(\phi\) at \(x(x_x,x_y,x_z)\) is:
\begin{align}
\label{equ:taylorExpansionLeaned}
\phi(p)&=\phi(x)+\nabla\phi(x)\cdot(p-x)+\frac12(p-x)^\top\nabla^2\phi(\xi)(p-x)\nonumber\\
&=\frac13(2p_x-2x_x+2p_y-2x_y+p_z-x_z)+R_2(p)\nonumber\\
&=\frac{g(p_x,p_y,p_z)}3+R_2(p),
\end{align}
where \(\xi=x+\lambda(p-x),\lambda\in[0,1]\).

Using the same continuous bound \(M\) for the Hessian as established in the \(T_\text{baseline}\) proof, it follows that \(\lvert R_2(p)\rvert\leq\frac12M\lVert p-x\rVert_2^2\leq\frac12MC^2h^2\).
Since \(g(p)\) is linear, its extreme values on the triangle \(f\) occur at its vertices. Assuming they occur at \(p_0\) and \(p_0+\Delta p(\Delta p_x,\Delta p_y,\Delta p_z)\), it can be obtained that
\begin{align}
\max(\lvert g(p_0)\rvert,\lvert g(p_0+\Delta p)\rvert)\nonumber&\geq\frac12(\lvert g(p_0)\rvert+\lvert g(p_0+\Delta p)\rvert)\nonumber\\
&\geq\frac12\lvert g(p_0+\Delta p)-g(p_0)\rvert\nonumber\\
&=\lvert\Delta p_x+\Delta p_y+\frac{\Delta p_z}2\rvert.\nonumber
\end{align}

To find the smallest \(\lvert\Delta p_x+\Delta p_y+\frac{\Delta p_z}2\rvert\), the 12 midpoints: \((ih+\frac h2,jh,kh)\), \((ih+h,jh+\frac h2,kh)\), \((ih+\frac h2,jh+h,kh)\), \((ih,jh+\frac h2,kh)\), \((ih,jh,kh+\frac h2)\), \((ih+h,jh,kh+\frac h2)\), \((ih+h,jh+h,kh+\frac h2)\), \((ih,jh+h,kh+\frac h2)\), \((ih+\frac h2,jh,kh+h)\), \((ih+h,jh+\frac h2,kh+h)\), \((ih+\frac h2,jh+h,kh+h)\), and \((ih,jh+\frac h2,kh+h)\) are considered. It can be verified that taking three arbitrary points from these 12 midpoints, calculating \(\lvert\Delta p_x+\Delta p_y+\frac{\Delta p_z}2\rvert\) between each pair of them, and taking the maximum value, the minimum of this maximum value is \(\frac h4\). Therefore, there exists a point \(p\in f\) such that \(\lvert g(p)\rvert\geq\frac h4\). Combining this with Equation \eqref{equ:taylorExpansionLeaned}, it comes to \(\lvert\phi(p)\rvert\geq\frac{\lvert g(p)\rvert}3-\lvert R_2(p)\rvert\geq\frac h{12}-\frac12MC^2h^2\). Using Equation \eqref{equ:SDF}, the shortest distance from \(p\) to \(S_\text{geom}\) is \(\lvert\phi(p)\rvert\). Taking the supremum over all points \(y\in T_\text{midpoint}\) in Equation \eqref{equ:hausdorffDistance}, it holds that \(d(S_\text{geom},T_\text{midpoint})\geq\sup_{y \in T_\text{midpoint}}\inf_{x\in S_\text{geom}}\lVert x-y\rVert_2\geq\frac h{12}-\frac12MC^2h^2\). For a sufficiently small grid size \(h\), the linear term \(\frac1{12}h\) dominates the quadratic term \(\frac12MC^2h^2\).

With both the upper and the lower bounds of \(d(S_\text{geom},T_\text{midpoint})\) being linear as \(h\to0\), Equation \eqref{equ:midpointConvergenceRate} is proved.

\addvspace{1em}
Finally to prove Equation \eqref{equ:MCHexConvergenceRate}, it is necessary to show that both the upper and the lower bounds of \(d(S_\text{geom},T_\text{MCHex})\) are quadratic as \(h\to0\).

\addvspace{1em}
\noindent\textbf{(5) Upper bound of \(d(S_\text{geom},T_\text{MCHex})\)}

For any point \(y\in T_\text{MCHex}\), it lies on a face \(f\subset T_\text{MCHex}\) inside a grid cell \(h_1\in H\). The vertices of \(f\), denoted as \(p_1,p_2,p_3\), are exact intersection points on the edges of \(h_1\). Therefore, it holds that \(\phi(p_i)=0\) for all \(i\). Since \(y\in f\), it can be written as a convex combination of these three vertices: \(y=\sum_iw_ip_i\), with \(\sum_iw_i=1\) and \(w_i\geq0\). The first-order Taylor expansion of \(\phi\) at \(y\) evaluated at \(p_i\) is
\begin{align}
\phi(p_i)=\phi(y)+\nabla\phi(y)\cdot(p_i-y)+\frac12(p_i-y)^\top\nabla^2\phi(\xi_i)(p_i-y),\nonumber
\end{align}
where \(\xi_i=y+\lambda_i(p_i-y)\) for some \(\lambda_i\in[0,1]\). Multiplying by \(w_i\) and summing over all vertices gives
\begin{align}
\sum_iw_i\phi(p_i)&=\phi(y)\sum_iw_i+\nabla\phi(y)\cdot\sum_iw_i(p_i-y)\nonumber\\
&+\frac12\sum_iw_i(p_i-y)^\top\nabla^2\phi(\xi_i)(p_i-y).\nonumber
\end{align}
Since \(\phi(p_i)=0\), \(\sum_iw_i=1\), and \(\sum_iw_i(p_i-y)=\sum_iw_ip_i-y\sum_iw_i=0\), the equation simplifies to
\begin{align}
\phi(y)=-\frac12\sum_iw_i(p_i-y)^\top\nabla^2\phi(\xi_i)(p_i-y).\nonumber
\end{align}
Using the continuous bound \(\lVert\nabla^2\phi(\xi)\rVert_2\leq M\) for the Hessian as established in the \(T_\text{baseline}\) proof, and noting that all points lie within the same grid cell \(h_1\), one has \(\lVert p_i-y\rVert_2\leq\sqrt3h\). Thus
\begin{align}
\lvert\phi(y)\rvert\leq\frac12M(\sqrt3h)^2\sum_iw_i=\frac32Mh^2.\nonumber
\end{align}
Using Equation \eqref{equ:SDF}, one has \(\inf_{x\in S_\text{geom}}\lVert y-x\rVert_2=\lvert\phi(y)\rvert\leq\frac32Mh^2\). Taking the supremum gives \(\sup_{y\in T_\text{MCHex}}\inf_{x\in S_\text{geom}}\lVert x-y\rVert_2\leq\frac32Mh^2\).

Conversely, for any \(x\in S_\text{geom}\), consider a line segment along the normal direction at \(x\) defined by \(y(t)=x+t\nabla\phi(x)\) for \(t\in[-2h,2h]\). By the first-order Taylor expansion of \(\phi\) at \(x\) along the line \(y(t)\), it follows that
\begin{align}
\label{equ:taylorExpansionLine}
\phi(y(t))&=\phi(x)+t\nabla\phi(x)\cdot\nabla\phi(x)+\frac12t^2\nabla\phi(x)^\top\nabla^2\phi(\xi)\nabla\phi(x)\nonumber\\
&=t+R_2(t),
\end{align}
where \(\xi=x+\lambda t\nabla\phi(x)\) for \(\lambda\in[0,1]\), and the second-order remainder satisfies \(\lvert R_2(t)\rvert\leq\frac12Mt^2\) using the same continuous bound \(M\) for the Hessian as established in the \(T_\text{baseline}\) proof. When \(t=-2h\), one has \(\phi(y(-2h))\leq-2h+2Mh^2\). For any corner vertex \(v\) of the grid cell containing the point \(y(-2h)\), its distance to \(y(-2h)\) is at most \(\sqrt3h\). Therefore
\begin{align}
\phi(v)\leq\phi(y(-2h))+\sqrt3h\leq-2h+\sqrt3h+2Mh^2.\nonumber
\end{align}
When \(h<\frac{2-\sqrt3}{2M}\), \(\phi(v)<0\) holds for all eight corners of this cell. Thus, this cell is completely inside \(S_\text{geom}\). Conversely, when \(t=2h\), one has \(\phi(y(2h))\geq2h-2Mh^2\). Using triangle inequality again gives
\begin{align}
\phi(v)\geq\phi(y(2h))-\sqrt3h\geq2h-\sqrt3h-2Mh^2.\nonumber
\end{align}
Similarly, when \(h<\frac{2-\sqrt3}{2M}\), \(\phi(v)>0\) holds for all eight corners of this cell.

Since the reconstructed mesh \(T_\text{MCHex}\) separates the grid vertices with negative values from those with positive values, the continuous line segment \(y(t)\) connecting a point in a completely inside cell to a point in a completely outside cell must intersect \(T_\text{MCHex}\) at some point \(y(t')\in T_\text{MCHex}\), where \(\lvert t'\rvert\leq2h\). From the upper bound of \(d(S_\text{geom},T_\text{MCHex})\) proven above, since \(y\in T_\text{MCHex}\), its signed distance satisfies \(\lvert\phi(y(t'))\rvert\leq\frac32Mh^2\).

On the other hand, substituting \(y(t')\) into the Taylor expansion Equation \eqref{equ:taylorExpansionLine} yields \(\phi(y(t'))=t'+R_2(t')\). Therefore
\begin{align}
\lvert t'\rvert\leq\lvert\phi(y(t'))\rvert+\lvert R_2(t')\rvert&\leq\frac32Mh^2+\frac12Mt'^2,\nonumber\\
\lvert t'\rvert(1-\frac12M\lvert t'\rvert)&\leq\frac32Mh^2,\nonumber\\
\lvert t'\rvert&\leq\frac{3Mh^2}{2(1-Mh)},\nonumber\\
\lVert x-y(t')\rVert_2=\lvert t'\rvert&<\sqrt3Mh^2.\nonumber
\end{align}
Taking the supremum over all \(x\in S_\text{geom}\), \(\sup_{x\in S_\text{geom}}\inf_{y\in T_\text{MCHex}}\lVert x-y\rVert_2<\sqrt3Mh^2\) is obtained. Combining both directions, the HD in Equation \eqref{equ:hausdorffDistance} satisfies \(d(S_\text{geom},T_\text{MCHex})\leq\sqrt3Mh^2\). In other words, the upper bound of \(d(S_\text{geom},T_\text{MCHex})\) is quadratic.

\addvspace{1em}
\noindent \textbf{(6) Lower bound of \(d(S_\text{geom},T_\text{MCHex})\)}

Since \(S_\text{geom}\) is a closed \(C^2\) compact surface, it has at least one convex point \(x_0\in S_\text{geom}\) where the Hessian \(\nabla^2\phi(x_0)\) restricted to the tangent plane is positive definite with its minimum eigenvalue \(m>0\). From the upper bound proven above, it holds that \(\inf_{y\in T_\text{MCHex}}\lVert x_0-y\rVert_2\leq\sqrt3Mh^2\). Let \(y_0\in T_\text{MCHex}\) be a point that achieves this infimum, which means \(\lVert x_0-y_0\rVert_2\leq\sqrt3Mh^2\). The point \(y_0\) must lie on a face \(f\subset T_\text{MCHex}\) generated within a grid cell \(h_1\). The vertices of \(f\), denoted as \(p_1,p_2,p_3\), are exact intersection points on the edges of \(h_1\), which means \(\phi(p_i)=0\) for all \(i\).

Any point \(p\in f\) has its distance to \(x_0\) bounded by \(\lVert p-x_0\rVert_2\leq\lVert p-y_0\rVert_2+\lVert y_0-x_0\rVert_2\leq\sqrt3h+\sqrt3Mh^2\). Now let \(p=\sum_{i=1}^3w_ip_i\) be the center of the face \(f\), where \(w_1=w_2=w_3=\frac13\). The first-order Taylor expansion of \(\phi\) at \(x_0\) evaluated at \(p\) is:
\begin{align}
\label{equ:taylorExpansionFaceCenter}
\phi(p)=\phi(x_0)+\nabla\phi(x_0)\cdot(p-x_0)+\frac12(p-x_0)^\top\nabla^2\phi(\xi)(p-x_0),
\end{align}
where \(\phi(x_0)=0\) and \(\xi=x_0+\lambda(p-x_0),\lambda\in[0,1]\). Evaluating Equation \eqref{equ:taylorExpansionFaceCenter} at each point \(p_i\) where \(\phi(p_i)=0\) yields
\begin{align}
\label{equ:taylorExpansionpi}
\nabla\phi(x_0)\cdot(p_i-x_0)&=-\frac12(p_i-x_0)^\top\nabla^2\phi(\xi_i)(p_i-x_0),\\
\sum_iw_i\nabla\phi(x_0)\cdot(p_i-x_0)&=\nabla\phi(x_0)\cdot(p-x_0)\nonumber\\
&=-\frac12\sum_iw_i(p_i-x_0)^\top\nabla^2\phi(\xi_i)(p_i-x_0),\nonumber
\end{align}
where \(\xi_i=x_0+\lambda_i(p_i-x_0),\lambda_i\in[0,1]\). Substituting this back into Equation \eqref{equ:taylorExpansionFaceCenter} leads to
\begin{align}
\label{equ:expandedPhi}
\phi(p)&=-\frac12\sum_iw_i(p_i-x_0)^\top\nabla^2\phi(\xi_i)(p_i-x_0)\nonumber\\
&+\frac12(p-x_0)^\top\nabla^2\phi(\xi)(p-x_0)\nonumber\\
&=-\frac12\sum_iw_i(p_i-p)^\top\nabla^2\phi(\xi_i)(p_i-p)\nonumber\\&-\sum_iw_i(p_i-p)^\top\nabla^2\phi(\xi_i)(p-x_0)\nonumber\\&+\frac12(p-x_0)^\top(\nabla^2\phi(\xi)-\sum_iw_i\nabla^2\phi(\xi_i))(p-x_0).
\end{align}
To bound the remainder terms, let \(\omega(\cdot)\) be the modulus of continuity of the Hessian \(\nabla^2\phi\) within the compact tubular neighborhood of \(S_\text{geom}\). Since \(\nabla^2\phi\) is continuous, one has \(\omega(\delta)\to0\) as \(\delta\to0\). Since \(\xi_i\) and \(\xi\) lie on the line segments connecting \(x_0\) to \(p_i\) and \(p\) respectively, \(\lVert\nabla^2\phi(\xi_i)-\nabla^2\phi(x_0)\rVert_2\leq\omega(\sqrt3h+\sqrt3Mh^2)\) and \(\lVert\nabla^2\phi(\xi)-\nabla^2\phi(x_0)\rVert_2\leq\omega(\sqrt3h+\sqrt3Mh^2)\) hold.

Let \(v_i=p_i-p\). \(v_i\) can be decomposed into tangential and normal components \(v_i=v_{i,\parallel}+v_{i,\perp}\nabla\phi(x_0)\), where \(v_{i,\perp}=v_i\cdot\nabla\phi(x_0)\). Since \(\lVert\nabla\phi(x_0)\rVert_2=\nabla\phi(x_0)^\top\nabla\phi(x_0)=1\), differentiating this yields
\begin{align}
\nabla(\nabla\phi(x_0)^\top\nabla\phi(x_0))=2\nabla^2\phi(x_0)\nabla\phi(x_0)=0,\nonumber
\end{align}
therefore
\begin{align}
v_i^\top\nabla^2\phi(x_0)v_i=v_{i,\parallel}^\top\nabla^2\phi(x_0)v_{i,\parallel}\geq m\lVert v_{i,\parallel}\rVert_2^2=m(\lVert v_i\rVert_2^2-v_{i,\perp}^2).\nonumber
\end{align}
To bound \(v_{i,\perp}\), taking the absolute value on Equation \eqref{equ:taylorExpansionpi} and applying the Hessian bound \(M\):
\begin{align}
\lvert\nabla\phi(x_0)\cdot(p-x_0)\rvert&\leq\frac13\sum_i\lvert\nabla\phi(x_0)\cdot(p_i-x_0)\rvert\nonumber\\
&\leq\frac13\sum_i\frac12M\lVert p_i-x_0\rVert_2^2\nonumber\\
&\leq\frac12M(\sqrt3h+\sqrt3Mh^2)^2.\nonumber
\end{align}
By the triangle inequality:
\begin{align}
\lvert v_{i,\perp}\rvert&=\lvert\nabla\phi(x_0)\cdot(p_i-p)\rvert\nonumber\\
&\leq\lvert\nabla\phi(x_0)\cdot(p_i-x_0)\rvert+\lvert\nabla\phi(x_0)\cdot(p-x_0)\rvert\nonumber\\
&\leq M(\sqrt3h+\sqrt3Mh^2)^2.\nonumber
\end{align}
The bound \(v_{i,\perp}^2\leq M^2(\sqrt3h+\sqrt3Mh^2)^4\) is obtained. With this bound, the first term in Equation \eqref{equ:expandedPhi} can be bounded by
\begin{align}
&\lvert\frac12\sum_iw_iv_i^\top\nabla^2\phi(\xi_i)v_i\rvert\nonumber\\
=&\frac16\sum_i\lvert v_i^\top\nabla^2\phi(x_0)v_i+v_i^\top(\nabla^2\phi(\xi_i)-\nabla^2\phi(x_0))v_i\rvert\nonumber\\
=&\frac16\sum_i\lvert v_{i,\parallel}^\top\nabla^2\phi(x_0)v_{i,\parallel}+2v_{i,\perp}v_{i,\parallel}^\top\nabla^2\phi(x_0)\nabla\phi(x_0)\nonumber\\
+&v_{i,\perp}^2\nabla\phi(x_0)^\top\nabla^2\phi(x_0)\nabla\phi(x_0)+v_i^\top(\nabla^2\phi(\xi_i)-\nabla^2\phi(x_0))v_i\rvert\nonumber\\
\geq&\frac16\sum_i(m\lVert v_{i,\parallel}\rVert_2^2+v_i^\top(\nabla^2\phi(\xi_i)-\nabla^2\phi(x_0))v_i)\nonumber\\
\geq&\frac16\sum_i(m\lVert v_i\rVert_2^2-mv_{i,\perp}^2-\omega(\sqrt3h+\sqrt3Mh^2)\lVert v_i\rVert_2^2)\nonumber\\
=&\frac16\sum_i((m-\omega(\sqrt3h+\sqrt3Mh^2))\lVert v_i\rVert_2^2-mM^2(\sqrt3h+\sqrt3Mh^2)^4).\nonumber\\
=&\frac1{18}(m-\omega(\sqrt3h+\sqrt3Mh^2))(\lVert p_1-p_2\rVert_2^2+\lVert p_2-p_3\rVert_2^2\nonumber\\
+&\lVert p_3-p_1\rVert_2^2)-\frac12mM^2(\sqrt3h+\sqrt3Mh^2)^4.\nonumber
\end{align}

The second term in Equation \eqref{equ:expandedPhi} can be bounded by
\begin{align}
&\lvert\sum_iw_i(p_i-p)^\top\nabla^2\phi(\xi_i)(p-x_0)\rvert\nonumber\\
=&\lvert\sum_iw_i(p_i-p)^\top(\nabla^2\phi(\xi_i)-\nabla^2\phi(x_0))(p-x_0)\rvert\nonumber\\
\leq&\sum_iw_i\lVert p_i-p\rVert_2\lVert\nabla^2\phi(\xi_i)-\nabla^2\phi(x_0)\rVert_2\lVert p-x_0\rVert_2\nonumber\\
\leq&\sqrt{3}h\omega(\sqrt3h+\sqrt3Mh^2)(\sqrt3h+\sqrt3Mh^2).\nonumber
\end{align}

The third term in Equation \eqref{equ:expandedPhi} can be bounded by
\begin{align}
&\lvert\frac12(p-x_0)^\top(\nabla^2\phi(\xi)-\sum_iw_i\nabla^2\phi(\xi_i))(p-x_0)\rvert\nonumber\\
\leq&\frac12(\sqrt3h+\sqrt3Mh^2)^2\lVert\nabla^2\phi(\xi)-\sum_iw_i\nabla^2\phi(\xi_i)\rVert_2\nonumber\\
\leq&\frac12(\sqrt3h+\sqrt3Mh^2)^2(\lVert\nabla^2\phi(\xi)-\nabla^2\phi(x_0)\rVert_2\nonumber\\
+&\sum_iw_i\lVert\nabla^2\phi(x_0)-\nabla^2\phi(\xi_i)\rVert_2)\nonumber\\
\leq&\frac12(\sqrt3h+\sqrt3Mh^2)^22\omega(\sqrt3h+\sqrt3Mh^2).\nonumber
\end{align}
Substituting these three bounds into Equation \eqref{equ:expandedPhi}, one has
\begin{align}
&\lvert\phi(p)\rvert\nonumber\\
\geq&\frac1{18}(m-\omega(\sqrt3h+\sqrt3Mh^2))(\lVert p_1-p_2\rVert_2^2+\lVert p_2-p_3\rVert_2^2\nonumber\\
+&\lVert p_3-p_1\rVert_2^2)-\frac12mM^2(\sqrt3h+\sqrt3Mh^2)^4\nonumber\\
-&\omega(\sqrt3h+\sqrt3Mh^2)(\sqrt3h+\sqrt3Mh^2)(2\sqrt3h+\sqrt3Mh^2)\nonumber\\
\geq&\frac1{18}(m-\omega(\sqrt3h+\sqrt3Mh^2))3(\frac{\sqrt2h}{1000})^2-\frac12mM^2(\sqrt3h+\sqrt3Mh^2)^4\nonumber\\
-&\omega(\sqrt3h+\sqrt3Mh^2)(\sqrt3h+\sqrt3Mh^2)(2\sqrt3h+\sqrt3Mh^2).\nonumber
\end{align}
Note that the last inequality comes from limiting the distance between \(p_i\) and the endpoints of its containing edge to be no less than \(\frac h{1000}\) to prevent degenerate edges, as introduced in Section \ref{sec:positiveJacobianProof}.

By Equation \eqref{equ:SDF}, the shortest distance from the center point \(p\in T_\text{MCHex}\) to \(S_\text{geom}\) is \(\lvert\phi(p)\rvert\). Taking the supremum over all points \(y\in T_\text{MCHex}\) in Equation \eqref{equ:hausdorffDistance}, one has
\begin{align}
&d(S_\text{geom},T_\text{MCHex})\nonumber\\
\geq&\sup_{y\in T_\text{MCHex}}\inf_{x\in S_\text{geom}}\lVert x-y\rVert_2\nonumber\\
\geq&\frac1{3000000}mh^2-\frac1{3000000}\omega(\sqrt3h+\sqrt3Mh^2)h^2\nonumber\\
-&\frac92mM^2(1+Mh)^4h^4-3\omega(\sqrt3h+\sqrt3Mh^2)(2+Mh)(1+Mh)h^2.\nonumber
\end{align}
For a sufficiently small grid size \(h\), \(\omega(\sqrt3h+\sqrt3Mh^2)\to0\), the quadratic term \(\frac1{3000000}mh^2\) dominates the higher-order terms of \(h\). 

With both the upper and the lower bounds of \(d(S_\text{geom},T_\text{MCHex})\) being quadratic as \(h\to0\), Equation \eqref{equ:MCHexConvergenceRate} is proved.

\end{proof}

\section{All-Hex Elements via Midpoint Subdivision: Sufficiency and Necessity}
\label{apd:only3RegularPolyhedronAfterMidpointSubdivisionGivesAllHex}

As shown in Figure \ref{fig:midpointSubdivisionExample}, elements in the complex \(P'\) are formed by generating a unique element, denoted \(c_v\), around each vertex \(v \in V\) of the original complex \(P\). Let \(\text{val}(v)\) be the face valence of vertex \(v\). The number of vertices, edges and faces of element \(c_v\) are checked as follows.

The vertices of element \(c_v\), denoted \(V_v\), are composed of: the original vertex \(v\); a new vertex \(v_P\) at the volume center of \(P\); the set of new vertices \(\{v_{f_i}\}_{i=1}^{\text{val}(v)}\) at the centers of faces adjacent to \(v\); and the set of new vertices \(\{v_{e_i}\}_{i=1}^{\text{val}(v)}\) at the midpoints of edges incident to \(v\). Therefore, the total number of vertices in \(c_v\) is \(\lvert V_v \rvert = 1\) (original vertex \(v\))\({}+1\) (volume center)\({}+\text{val}(v)\) (face centers)\({}+\text{val}(v)\) (edge midpoints) \(=2\text{val}(v)+2\).

The edges of element \(c_v\), denoted \(E_v\), are composed of: edges connecting the original vertex \(v\) to the midpoints of its incident edges \(\{v_{e_i}\}_{i=1}^{\text{val}(v)}\); edges connecting the volume center \(v_P\) to the centers of the adjacent faces \(\{v_{f_i}\}_{i=1}^{\text{val}(v)}\); edges connecting each face center \(v_{f_i}\) to the midpoints of its two edges \(v_{e_i}\) and \(v_{e_{i+1}}\) (with indices taken modulo \(\text{val}(v)\)). Therefore, the total number of edges in \(c_v\) is \(\lvert E_v\rvert=\text{val}(v)+\text{val}(v)+2\text{val}(v)=4\text{val}(v)\).

The faces of element \(c_v\), denoted \(F_v\), are composed of two categories: quads formed by the original vertex \(v\), a face center \(v_{f_i}\), and two edge midpoints \(v_{e_i}\) and \(v_{e_{i+1}}\); quads formed by the volume center \(v_P\), an edge midpoint \(v_{e_{i+1}}\), and two face centers \(v_{f_i}\) and \(v_{f_{i+1}}\) adjacent to that edge. Each category has \(\text{val}(v)\) faces. Therefore, the total number of faces in \(c_v\) is \(\lvert F_v\rvert=2\text{val}(v)\).

To prove the if and only if statement, it is necessary to prove both the \(\rightarrow\) direction and the \(\leftarrow\) direction.

(\(\rightarrow\)) Assume that every element \(c\in C'\) is a hex. This means that for any given vertex \(v \in V\), the corresponding element \(c_v\) is a hex. By definition, a hex is an element with eight vertices. Thus, for any \(c_v\), \(\lvert V_v \rvert = 8\). Using the established vertex count formula, it follows that \(2\text{val}(v) + 2 = 8\), therefore \(\text{val}(v)=3\). Since this condition must hold for every vertex \(v \in V\), the polyhedron \(P\) must be 3-regular.

(\(\leftarrow\)) Assume that \(P\) is a 3-regular polyhedron. This means that for every vertex \(v \in V\), its valence is \(\text{val}(v) = 3\). Therefore, for element \(c_v\), \(\lvert V_v \rvert = 2\text{val}(v) + 2 = 2(3) + 2 = 8\), \(\lvert E_v \rvert = 4\text{val}(v) = 4(3) = 12\), \(\lvert F_v \rvert = 2\text{val}(v) = 2(3) = 6\). An element with eight vertices, 12 edges, and six quad faces must be a hex. Therefore, if \(P\) is 3-regular, every element \(c_v\) is a hex.

Since both the sufficiency and necessity conditions hold, it is proved that every element \(c\in C'\) is a hex if and only if \(P\) is a 3-regular polyhedron.

\section{Conversion from Monomial to Bernstein Coefficients}
\label{apd:monomialCoefficientsToBernsteinCoefficients}

Using the factorial definition of combinatorial numbers, it holds that:
\begin{align}
\label{equ:combinatorial}
\binom{n_j-k_j}{i_j-k_j}=\frac{(n_j-k_j)!}{(i_j-k_j)!(n_j-i_j)!}
=\frac{\binom{i_j}{k_j}\binom{n_j}{i_j}}{\binom{n_j}{k_j}}.
\end{align}
Using the binomial theorem and Equation \eqref{equ:combinatorial}, it holds that:
\begin{align}
\label{equ:xjkj}
x_j^{k_j}&=x_j^{k_j}(x_j+(1-x_j))^{n_j-k_j}\nonumber\\
&=x_j^{k_j}\sum_{i_j=0}^{n_j-k_j}\binom{n_j-k_j}{i_j}{x_j}^{i_j}(1-x_j)^{n_j-k_j-i_j}\nonumber\\
&=\sum_{i_j=0}^{n_j-k_j}\binom{n_j-k_j}{i_j}{x_j}^{i_j+k_j}(1-x_j)^{n_j-(i_j+k_j)}\nonumber\\
&=\sum_{i_j=k_j}^{n_j}\binom{n_j-k_j}{i_j-k_j}{x_j}^{i_j}(1-x_j)^{n_j-i_j}\nonumber\\
&=\sum_{i_j=k_j}^{n_j}\frac{\binom{i_j}{k_j}\binom{n_j}{i_j}}{\binom{n_j}{k_j}}{x_j}^{i_j}(1-x_j)^{n_j-i_j}.
\end{align}

Therefore, beginning with the polynomial expansion of \(\mathbf{J}(\mathbf{x})\), by substituting the expansion for each \(x_j^{k_j}\) using Equation \eqref{equ:xjkj}, interchanging the product and sum, and changing the summation order to factor out the basis functions \(B_{\mathbf{i},\mathbf{n}}(\mathbf{x})\), it can be obtained
\begin{align}
\mathbf{J}(\mathbf{x})&=\sum_{\mathbf{0}\leq\mathbf{k}\leq\mathbf{n}}J_{\mathbf{k}}\prod_{j=1}^mx_j^{k_j}\nonumber\\
&=\sum_{\mathbf{0}\leq\mathbf{k}\leq\mathbf{n}}J_{\mathbf{k}}\prod_{j=1}^m(\sum_{i_j=k_j}^{n_j}\frac{\binom{i_j}{k_j}}{\binom{n_j}{k_j}}\binom{n_j}{i_j}x_j^{i_j}(1-x_j)^{n_j-i_j})\nonumber\\
&=\sum_{\mathbf{0}\leq\mathbf{k}\leq\mathbf{n}}J_{\mathbf{k}}\sum_{\mathbf{k}\leq\mathbf{i}\leq\mathbf{n}}\prod_{j=1}^m\frac{\binom{i_j}{k_j}}{\binom{n_j}{k_j}}\binom{n_j}{i_j}x_j^{i_j}(1-x_j)^{n_j-i_j}\nonumber\\
&=\sum_{\mathbf{0}\leq\mathbf{i}\leq\mathbf{n}}(\sum_{\mathbf{0}\leq\mathbf{k}\leq\mathbf{i}}\prod_{j=1}^m\frac{\binom{i_j}{k_j}}{\binom{n_j}{k_j}}J_{\mathbf{k}})\prod_{j=1}^m\binom{n_j}{i_j}x_j^{i_j}(1-x_j)^{n_j-i_j}\nonumber\\
&=\sum_{\mathbf{0}\leq\mathbf{i}\leq\mathbf{n}}\beta_{\mathbf{i}}B_{\mathbf{i},\mathbf{n}}(\mathbf{x}).
\end{align}

\noindent Therefore, Equation \eqref{equ:bernsteinCoefficient} holds and the proof is done. This proof extends the univariate case established in \cite{mathar2018orthogonal}, Appendix B, to the multivariate setting. This shows that the coefficients \(\beta_\mathbf{i}\) of the Bernstein basis are related to the monomial coefficients \(J_\mathbf{i}\) via a linear transformation.
\balance

\end{document}